\documentclass[nonumbers]{aastex631}

\usepackage{etoolbox}%

\newcommand{\sgra}{Sgr A$^{\star}$}

\usepackage{graphicx}

\usepackage{CJKutf8}
\usepackage{bm}
\usepackage{natbib}
\usepackage{amsmath}
\usepackage{booktabs}
\usepackage{threeparttable}
\usepackage{hyperref} 
\usepackage{etoolbox}

\begin{document}
\begin{CJK*}{UTF8}{gbsn}

\title{The complex kinematics of the young stars orbiting the supermassive black hole in the Galactic center can be explained by  the presence of an intermediate mass companion of Sgr A$^\star$}

\author[0000-0002-7814-9185]{Xiaochen Zheng（郑晓晨)}
\affiliation{Beijing Planetarium, Beijing Academy of Science and Technology \\
No. 138 Xizhimenwai Main Street, Beijing 100044, China}

\author{Long Wang (王龙)}
\affiliation{School of Physics and Astronom, Sun Yat-sen University \\
Daxue Road, Zhuhai 519082, China}
\affiliation{CSST Science Center for the Guangdong-Hong Kong-Macau Greater Bay Area \\
Zhuhai 519082, China}

\author{Douglas N. C. Lin（林潮）}
\affiliation{Department of Astronomy and Astrophysics, University of California Santa Cruz \\
CA 95064, USA}
\affiliation{Institute for Advanced Studies, Tsinghua University \\
Beijing 100084, China}

\author{Andreas Burkert}
\affiliation{Universitats-Sternwarte Munchen, Ludwig-Maximilians University Munich \\
University Observatory Scheinerstr. 1, Munchen D-81679, Germany}
\affiliation{Max-Planck Institute for Extraterrestrial Physics \\
Giessenbachstr. 1, Garching D-85748, Germany}

\author{Shude Mao （毛淑德）}
\affiliation{Department of Astronomy, Westlake University\\
Hangzhou 310024, China}

\correspondingauthor{Xiaochen Zheng, Long Wang, Douglas N. C. Lin, Andreas Burkert, Shude Mao}
\email{x.c.zheng1989@gmail.com, 
wanglong8@mail.sysu.edu.cn,
lin@ucolick.org,
burkert@usm.lmu.de,
shude.mao@westlake.edu.cn}

\begin{abstract}
The sub-parsec proximity around the Sgr A$^\star$ supermassive black hole (SMBH) 
in the center of the Milky Way contains an inner cluster of eccentric S-stars with randomly 
oriented orbits, a midway-disk of clockwise-rotating stars (CWSs), and a surrounding population 
of off-the-disk stars (ODSs).  Despite their diverse kinematic properties, all 
three-populations appear to be massive (WR/O/B types) and have similarly limited 
life span $\tau_\star \sim 6-15$ Myr. Several scenarios, including star formation induced by
SMBH's close encounters with one or more gas clouds as well as impulsive close 
scattering by a putative intermediate-mass companion (IMC) of Sgr A$^\star$ possible 
an intermediate-mass black hole (IMBH), have been proposed to  
explain piecemeal for the origin and dynamical evolution of S-stars, CWSs, ODSs, as 
well as hyper-velocity stars in the Galaxy. But, their coexistence and the origin of a 
recently discovered zone of avoidance in S-stars' eccentricity-peri-centric-distance
distribution remain enigmatic.  Here, we construct a unified model to comprehensively
take into account these stars' interaction with each other, their single natal disk, and an 
independent IMC. We show their disparate present-day orbits would only be concurrently attainable,
within their multi-Myr age, under the combined influence of IMC's secular perturbation 
and these stars' resonant relaxation in a depleting gaseous-disk environment.
\end{abstract}


%%%%%%%%%%%%%%%%%%%%%%%%%%%%%%%%%%%%%%%%%%%%%
\section{Introduction} \label{sec:intro}

With a mass $M_\bullet = 4  \times 10^6 M_\odot$
\citep{Genzel1997, Ghez1998, schodel2002, ghez2005, eisenhauer2005, gravity2023-2} and
gravitational radius $r_\bullet=G M_\bullet/c^2 = 0.02$ AU,
Sgr A$^\ast$ is surrounded by a complex configuration of young massive stars. 
At least 195 young stars co-exist within $\sim 0.5$ pc from \sgra at the Galactic center \citep{vonfellenberg2022}, they belong to multiple young-star populations, including: 

\noindent
1) a group of $\sim 40$ known B stars with mass $\sim 8-20~M_\odot$ and nearly isotropic velocity dispersion, the relatively small semi-major axis
($a_\star \sim 5\times 10^{-3}-0.04$ pc), short orbital periods ($P_\star =2 \pi \sqrt{a_\star^3 /GM_\bullet} \geq 16$ yr), high eccentricity ($e_\star \sim 0.4-0.97$), and nearly random inclinations \citep{%Genzel2003, Ghez2003a, 
gillessen2017}, which are commonly referred to as the S-stars \citep{%Genzel2003, Ghez2003a, 
gillessen2017}. Recent observations 
\citep{burkert2024} show a non-uniform eccentricity-pericenter distance $r_p=
(1-e_\star)a_\star$ distribution with a zone of avoidance \citep{Generozov2025};

\noindent
2) about $\sim 20 \%$ of the entire young-star population belong to a subgroup of  
O/WR stars with top-heavy luminosity function reside in a possibly-warped 
disk with clock-wise rotation, $a_{\star} \sim 0.04-0.5$ pc,  
$e_\star \sim 0.2-0.4$, and they are commonly referred to as the clockwise-rotating stellar disk 
stars (CWSs) \citep{levin2003, paumard2006, lu2006, lu2009, lockmann2009, bartko2009, bartko2010, yelda2014};

\noindent 
3) while most of the CWSs orbit in the same direction with a limited range of
inclinations relative to an average plane, the large remaining fraction of the young stars are the ``off-disk'' stars (ODSs) with much less well-determined, probably large $e_\star$ and inclination
$i_\star$ but non-isotropic distribution, surrounding the disk-star population.
Several over-dense concentrations in their kinematic distribution have been 
identified, including a controversial group of counter-clockwise
rotating stars \citep{paumard2006, lu2009, yelda2014, ali2020}.  It is not clear to what extent are the apparent ``patchiness'' due to the incompleteness of the kinematic data. 
The luminosity function of the ODSs
is similar to that of the S-stars and less top-heavy than the disk stars 
\citep{vonfellenberg2022}.

In addition, there are $\sim 10^7$ less massive, mature stars with  
$a_\star \sim 1-2$ pc, which are commonly referred to as the nuclear cluster stars
\citep{Do2009, schodel2009, schodel2014, Genzel2010}.  

Interestingly, the S-stars and the ODSs are of B-type with an upper age limit of $\leq 15$ Myr \citep{gillessen2017, habibi2017}. In addition to B-type \citep{levin2003, paumard2006},
the CWSs are mostly O/WR type with ages $\tau_\star \sim 6 \pm 2$ Myr \citep{ghez2003b, 
vonfellenberg2022}. These dichotomies suggest S-stars, CWSs, and ODSs have 
undergone disparate evolutionary paths with diverse kinematic outcomes. %even though they originated from a common natal disk.

The origin and dynamical evolution of the young stellar populations near the Galactic center, the S-stars, the Clockwise Disk stars, and the Off-disk stars are typically studied in isolation. A leading hypothesis for the CWS stars is that they formed {\it in situ} within a gaseous disk around Sgr A$^\star$ \citep{Goodman2003, levin2003, BonnellRice2008, HobbsNayakshin2009}. Their current, modest eccentricities and limited inclination range could then be excited by their mutual gravitational interactions and corresponding gravitational resonance effects \citep{rauch1996, Alexander2007, Madigan2011, rantala2024}.

However, this mechanism of local relaxation is insufficient to explain the entire stellar population. It cannot produce the observed S-stars and ODSs, which exhibit high eccentricities and inclinations, within their short lifetimes ($\tau_{\star} \lesssim 10$ Myr) \citep{Freitag2006}. Alternative formation channels of CWSs have been proposed, including the tidal disruption of an infalling young star cluster \citep{gerhard2001, Genzel2003, Kim2003, maillard2004, Levin2005}. Other models invoke external perturbers to explain the CWSs' warp and hyper-velocity stars, such as a stellar cluster \citep{schodel2005}, a second, inclined disk (e.g., counter-clockwise disk) \citep{paumard2006, lockmann2009}, or an inspiraling intermediate-mass black hole (IMBH) \citep{yu2007}. Yet, many of these scenarios struggle to reconcile the S-stars' high central concentration with observational constraints, which place stringent upper limits on the mass and lower limits on the semi-major axis of any central intermediate-mass companion  \citep{gravity2023}. 

A potential mechanism to jump-start the S-stars' large velocity dispersion is through the disruption of an inspiralling young stellar cluster with an IMBH at its center \citep{gerhard2001, Hansen2003, Gurkan2005, Merritt2009} or by scattering with an intermediate mass companion (not necessarily a black hole) of Sgr A$^\star$ \citep{Levin2005, levin2006, PZ2006, Matsubayashi2007, yu2007, lockmann2008, Gualandris2009, Gualandris2010}. 
The role of tidally disrupted binaries was originally proposed by 
\cite{Hills1988} and has been developed by subsequent works to explain both bound S-stars and hyper-velocity stars \citep[e.g.,][]{Perets2007, GenerozovMadigan2020, Generozov2020, Generozov2021, Generozov2022}. In the most recent work, \cite{Generozov2025} investigated how a binary approaches the SMBH within its tidal radius to produce the S-star cluster. They can naturally explain the recently discovered zone of avoidance by the binary tidal disruption physics combined with relaxation. However, in their work,  the progenitor binaries originate from large scales (5-100 pc) in the nuclear stellar cluster, with the initial eccentricity being extremely high ($e$ close to $1$). Such a hypothetical population of young (5-9 Myr old) binary stars is yet to be observationally identified.  Moreover, such a process is unlikely to also account for the coexistence of the nearby disk stars with low-eccentricity orbits.

In this work, we present a unified model to coherently address the present-day kinematic diversity among the S-stars, CWSs, and ODSs, by considering an alternative possibility that they formed {\it simultaneously}, {\it in situ}, in a {\it single} natal disk and evolved {\it concurrently}. 
We propose that these populations share a common origin, having formed simultaneously and in situ within a single natal disk. We then model their concurrent evolution, wherein their orbits are shaped by both their intrinsic mutual interactions and the long-range secular perturbation from an {\it distant} intermediate-mass component (IMC). 

The combined influences of the von Zeipel-Lidov-Kozai effect (\S\ref{sec:kozai})
and the sweeping secular resonance (\S\ref{sec:sweeping})
of the IMC which acts during the depletion of their natal disk, can efficiently excite stellar eccentricities and inclinations around the vicinity of the IMC and increase the angular momentum deficit of the stellar system 
(\S\ref{sec:dynrelax}). Some highly excited CWSs and ODSs have orbital perigee that can penetrate the S-stars domain region. Ensuing resonant relaxation redistributes the angular momentum between these high eccentricity intruders and indigenous stars, randomizing their eccentricities and orbital inclinations, building up the spheroidal S-star cluster. With a longer resonant relaxation timescale, the CWSs' residual disk is preserved with modest orbits and an outer warp as observed.
%scattering a subset of stars onto high-dispersion, S-star-like orbits.

This model synthesizes three key aspects—common origin, concurrent evolution, and distant secular perturbation—into a single framework for the first time.
We therefore further explore how a distant intermediate-mass perturber, through gravitational interactions, can excite eccentricity ($e_\star$) and inclination ($i_\star$), thereby efficiently populating the S-star region around Sgr A$^\star$. 
This hypothetical intermediate-mass component (IMC) can be the core of a captured stellar cluster \citep{wang2023, gerhard2001} and/or an intermediate-mass black hole \citep{PZ2006,Matsubayashi2007,yu2007,Hansen2003,Gurkan2005,Levin2005,baumgardt2006,levin2006,lockmann2008,Gualandris2009,Merritt2009,Gualandris2010, GenerozovMadigan2020}. The cluster IRS-13E, with an estimated mass ($M_{\rm IMC} = 10^4~M_\odot$), observed at projected separation 0.13 pc~\citep{Krabbe1995,
baganoff2003, maillard2004, tsuboi2017}, even though the existence of the IMBH remains controversial \citep{schodel2005, Fritz2010}, would be an excellent candidate \citep{kocsis2011}
and is compatible with the stringent upper limits placed on $M_{\rm IMC}$ very close to Sgr A$^\star$ \citep{gravity2023, gravity2023-2}.

We present simulations which reproduce S-stars' high eccentricities, inclinations, and radial distribution, including a zone of avoidance within their lifespan ($\tau_\star$). We also
account for the apparent coexistence of the CWSs and another stellar disk 
among the ODSs, without the assumption of separate natal clouds/clusters or 
substantial migration of young stars \citep{BonnellRice2008, HobbsNayakshin2009,  
alig2013, Perets2009, Perets2010,levin2024}.

In Section \ref{sec:method}, we introduce the dominant physical processes and detail our configurations for the numerical models, encompassing the orbits of IMC, the initial kinematic properties of young stellar populations, and the gaseous disk components. Section \ref{sec:results} presents an analysis of the outcomes from the \textit{Fiducial} model, alongside other models, and juxtaposes these findings with observational data. The concluding section is dedicated to summarizing our findings and engaging in a broader discussion of the implications and significance of our results.

%%%%%%%%%%%%%%%%%%%%%%%%%%%%%%%%%%%%%%%%%%%%%%
\section{Methods}
\label{sec:method}

To quantitatively infer the origin of this complex stellar configuration, we carry out
simulations with a universal disk of coeval seeds for the S-stars, CWSs, and ODSs. 
These progenitors were either formed spontaneously \citep{Goodman2003, Nayakshin2007, levin2007} or were captured and rejuvenated \citep{syer1991, artymowicz1993, davies2020} 
there several Myr ago.

%%%%%%%%%%%%%%%%%%%%%%%%%%%%%%%%%%%%%%%%%%%%%%%%%%%%%%%
\subsection{Computational methods and model parameters}
\label{sec:model}
%In this work, we mainly carry out statistical analysis on the orbital evolution (including semi-major axes $a_\star$, eccentricity $e_\star$, and inclination $i_\star$) of young stars formed in nearly circular, planar orbits (i.e. $e_\star\simeq i_\star \simeq 0$) with a range of initial semi-major axis ($a_0$) around the SMBH. 

We analyze the dynamical evolution of these young stars around the supermassive black holes (SMBH, with mass $M_\bullet$ and gravitational radius $r_\bullet$) under IMC's and their mutual gravitational influence. We adopt
an open-source \textit{N}-body code PeTar \citep{wang+2020b}. 
A fourth-order Hermite integrator \citep{aarseth2003} is used to integrate the orbits of the stars, IMC, and the central SMBH. The slow-down algorithmic-regularization method \citep{wang+2020a} is used for integrating close encounters between stars.

In these simulations, we neglect the relativistic correction and Lense-Thirring precession \citep{hopman2006, levin2007}, which may be important for stars undergoing very close encounters with Sgr A$^\star$ \citep{levin2003, iorio2020, peissker2020, fragione2020}, especially in the context of tidal disruption events and ejection of hyper-velocity stars \citep{zheng+2021}. 
The periastra of known stars around Sgr A$^\star$ are sufficiently far away to avoid tidal disruption 
\citep{Hills1975, frank1976}. Nevertheless, very few stars may venture into regions where
post-Newtonian corrections \citep{rodriguez2018}, including the Lense-Thirring effect
\citep{lense1918, merritt2013, iorio2020} become non-negligible \citep{zheng+2021}. However, 
they are unlikely to influence the dynamical evolution of the S-stars, CWSs, and ODSs \citep{Tomar2024}.
First and second-order post-Newtonian (pN) corrections are included in some test models ({\it pN} model in F, \S\ref{sec:complement}) but they are negligible for most S-stars, CWSs, and ODSs.

We have compiled a summary of all the models in this study, as presented in Table \ref{tab:models}. The models are categorized based on the settings of either {\bf Stars}, {\bf Gas disk} or {\bf IMC}. The {\it Fiducial} model is highlighted within a box.

\begin{table}[!ht]
\centering
\caption{Parameters for the Numerical Simulations}
\begin{threeparttable}
%\label{all_models}
%\begin{tabularx}{2\textwidth}{|X|X|X|X|X|X|X|}
%\resizebox{\textwidth}{!}
\begin{tabular}{c | c c c c c c c}
%{\bf Model ID}
\toprule
%\rule{1.5\textwidth}{1mm}
%\vspace{1mm}
%\rule{\textwidth}{1mm} \\
 {\bf Model id}& & &\underline{Stars} \\
& $N_{\star}$ \tnote{[1]} & $a_{0}$ (pc) \tnote{[2]} & $S_\star$ \tnote{[3]} & $e_{0}$ \tnote{[4]} & $m_{\star}$ $(M_{\odot})$ \tnote{[5]} & GR \tnote{[6]}\\
\midrule
\boxed{\rm Fiducial} & 500 & $0.01-0.2$ & -1.5 & C & 15 & $-$ \\
Extended Disk & 500 & $0.005-0.2$ & -1.5 & C & 15 & $-$\\
Steep & 500 & $0.01-0.2$ & -2 & C & 15 & $-$ \\
No CWSs & 310 & $0.01-0.1$ & -1.5 & C & 15 & $-$\\
Eccentric S-stars & 310 & $0.01-0.1$ & -1.5 & E & 15 & $-$ \\
Massless & 500 & $0.01-0.2$ & -1.5 & C & 0.1 & $-$ \\
pN & 500 & $0.01-0.2$ & -1.5 & C & 15 & $+$ \\
IMF & 500 & $0.01-0.2$ & -1.5 & C & Kroupa & $-$ \\
%IMF 2 & 500 & $0.01-0.2$ & -1.5 & C & Salpeter & $-$ \\
%Bi-modal & 1000 & $0.01-0.2$ & -2.5 & C & 15+5 & $-$ \\
\midrule
\midrule
{\bf Model id} & & & \underline{Gas Disk} \\
 & & $\Sigma_0$ $({\rm g/cm^2})$ \tnote{[7]} & & $\tau_{\rm dep}$ (Myr) \tnote{[8]} \\
 \midrule
\boxed{\rm Fiducial} & & 800 & & 2.5  \\
Massive Disk 0 & & 8000 & & 2.5   \\
Massive Disk 1 & & 8000 & & 1  \\
No Gas Disk & & 0 & & $-$  \\
\midrule
\midrule
{\bf Model id} & & & \underline{IMC} \\
& $i_{\rm IMC}^\prime$ $(^{\circ})$ \tnote{[9]} & $N_{\rm IMC}$ \tnote{[10]} & $M_{\rm IMC}$ $(M_{\odot})$ \tnote{[11]} & $\Delta \theta$ $(^{\circ})$ \tnote{[12]} \\
\midrule
\boxed{\rm Fiducial} & 120 & 1 & $10^4$ & $-$  \\
INC-0 &  0 & 1 & $10^4$ & $-$\\
INC-30 & 30 & 1 & $10^4$ & $-$\\
INC-60 &  60 & 1 & $10^4$ & $-$\\
INC-90 &  90 & 1 & $10^4$ & $-$\\
INC-150 &  150 & 1 & $10^4$ & $-$\\
INC-180 &  180 & 1 & $10^4$ & $-$\\
No IMC &  $-$  & $-$ & $-$ & $-$\\
2IMC-10 &  120  & 2 & $5 \times 10^3$ & 10 \\
2IMC-20 &  120  & 2 & $5 \times 10^3$ & 20\\
2IMC-60 &  120  & 2 & $5 \times 10^3$ & 60\\
2IMC-180 & 120  & 2 & $5 \times 10^3$ & 180\\
10IMC &  120  & 10 & $10^3$ & 18\\
%\midrule
%\midrule
%{\bf Model ID}                                                     
% & & \underline{Parameters of IMC} \\
% & IMC & $i_{\rm IMC}^\prime$ $(^{\circ})$ \footnotemark[8]  \\
% \midrule
%\boxed{Fiducial}  & W & 120 \\
%IMC-0 & W & 0  \\
%IMC-30 & W & 30  \\
%IMC-60 & W & 60  \\
%IMC-90 & W & 90 \\
%IMC-150 & W & 150  \\
%IMC-180 & W & 180 \\
%No IMC & N & -  \\
%\midrule
%{\bf Model ID}  & &  \underline{post-Newtonion} \\
%\midrule
%Fiducial & &  W \\
%pN & & N \\
\bottomrule
\end{tabular}
\begin{tablenotes}
%\small
\item[1] The number of stars in each run.
\item[2] The initial semi-major axis distribution of young stars, including the inner and outer boundaries.
\item[3] The power index of stars' initial surface density distribution.
\item[4] The initial eccentricity of young stars: C for circular orbits, and E for orbits with an eccentricity that follows a normal distribution ranging from 0.01 to 0.5, the initial angular momentum deficit AMD following the eccentric-stream scenario \citep{Alexander2007, BonnellRice2008, HobbsNayakshin2009}.
\item[5] The initial mass of young stars.
\item[6] Turn on($+$)/off($-$) the first and second-order post-Newtonian corrections.
\item[7] The initial surface density of the gas disk at $R_0$.
\item[8] The depletion timescale of the gas disk.
\item[9] The initial relative inclination between stars and IMC.
\item[10] The number of IMC.
\item[11] The mass of IMC.
\item[12] The initial relative face angle between two IMCs.
\end{tablenotes}
%}
%\footnotetext{}
%\footnotetext[0]{Number of finished runs.}
%\footnotetext[1]{The number of stars in each run.}
%\footnotetext[2]{The initial semi-major axis distribution of young stars, including the inner and outer boundaries.}
%\footnotetext[3]{The power index of stars' initial surface density distribution.}
%\footnotetext[4]{The initial eccentricity of young stars: C for circular orbits, and E for orbits with an eccentricity that follows a normal distribution ranging from 0.01 to 0.5, the initial angular momentum deficit AMD following the eccentric-stream scenario \citep{Alexander2007, bonnell2008, hobbs2009}.}
%with $e_0=0.5$ 
%\footnotetext[5]{The initial mass of young stars.}
%\footnotetext[6]{Turn on($+$)/off($-$) the first and second-order post-Newtonian corrections.}
%\footnotetext[7]{The initial surface density of the gas disk at $R_0$.}
%\footnotetext[8]{The depletion timescale of the gas disk.}
%\footnotetext[9]{The initial relative inclination between stars and IMC.}
%\footnotetext[10]{The number of IMC.}
%\footnotetext[11]{The mass of IMC.}
%\footnotetext[12]{The initial relative face angle between two IMCs.}
\end{threeparttable}
\label{tab:models}
\end{table}

%%%%%%%%%%%%%%%%%%%%%%%%%%%%%%%%%%%%%%%%%%%%%%%%%%%%%%
\subsubsection{IMC's orbit}

Our model in fact portrays a remarkably accurate analogue to the Solar System on Galactic scales. 
The SMBH represents our Sun, the IMC corresponds to Jupiter, the disk stars are equivalent to 
the asteroid belt objects. This work highlights IMC's secular distant perturbation on the orbital evolution of CWSs within a few 0.1 pc from the SMBH. This effect is
analogous to Jupiter's influence on asteroids in the main belt  
\citep{nagasawa+2005, zheng+2017a, zheng+2017b}, and this process has been analyzed and discussed in the context of the Galactic center \citep{zheng+2020, zheng+2021}.
Its celestial perturbation is equivalent to that imposed by a stream of co-orbital clusters \citep{murray2000}.  
And the S-stars are synonymous to the near-Earth asteroids that have been 
scattered out of the belt by gravitational interaction with Jupiter. 

Our simulations take into account stars' gravitational interaction among themselves and with an IMC.  There are some stringent upper mass limits on any IMC very close 
to the Sgr A$^\star$ (interior to the orbit of the S2 star) \citep{gravity2023}, 
albeit they do not exclude the possibility of distant IMCs.
A cluster, IRS 13E, with an estimated mass $M_{\rm IMC} = 10^4 M_\odot$ and at a projected distance 0.13 pc \citep{Krabbe1995, baganoff2003, maillard2004, tsuboi2017} remains a viable IMC candidate. 

Based on extensive model-parameter studies of these previous simulations \citep{zheng+2020, zheng+2021}, we select a {\it Fiducial} model for the IMC with an orbital semi-major axis $a_{\rm IMC} = 0.35$~pc, period $P_{\rm IMC} \sim 10^4$ yr, eccentricity $e_{\rm IMC} = 0.3$, inclination  $i_{\rm IMC}^\prime =i_{\rm IMC}-i_0= 120^{\circ}$ relative to the initial stellar and gaseous disk's $i_0$. Both $i_0$ and $i_{\rm IMC}$ are relative to the I-plane (plane perpendicular to the initial total angular-momentum vector).

Since an IMC is invoked in our model to provide distant secular perturbations rather than close encounters, it can be residual cores or moving fragments of infalling stellar clusters \citep{wang2023} or compact objects 
\citep{PZ2006,Matsubayashi2007,yu2007}. 

With the Lagrange-Laplace orbit-averaging approach \citep{murray2000}, a single IMC's secular perturbation is equivalent to multiple sub-systems with the same total mass and orbital elements (longitude of periapse $\varpi_{\rm IMC}$ and longitude of ascending node $\Omega_{\rm IMC}$) at random phases. These sub-systems may represent the tidally-disrupted relics of a parent star cluster. We simulate models {\it 2IMC-180} (\S\ref{sec:complement}G), {\it 2IMC-60}, {\it 2IMC-20}, {\it 2IMC-10} with two IMCs with equal mass $0.5\times10^4 M_{\odot}$ and various face angles ($\theta$), $\Delta \theta = 180^{\circ}, 60^{\circ}, 20^{\circ}$ and $10^{\circ}$, respectively. We also consider ten $10^3 M_{\odot}$ IMCs with equal initial phase separations on the same orbit in Model {\it 10IMC} (see details in Table \ref{tab:models}). 

We also examine the dependence of the results on the inclination of the IMC. This is investigated through a series of additional models, {\it INC-0}, {\it INC-30}, {\it INC-60}, {\it INC-90}, {\it INC-150} and {\it INC-180}, which correspond to inclinations of $i_{\rm IMC} ^\prime =0, \pi/6, \pi/3, \pi/2, 5 \pi/6,$ $\pi$, respectively (see details in Table \ref{tab:models}). 

%%%%%%%%%%%%%%%%%%%%%%%%%%%%%%%%%%%%%%%%%%%%%%%%%%%%%%%%%%%%
\subsubsection{Initial kinematic properties of disk stars}
\label{sec:diskstars}

Even though some S-stars, CWSs, and ODSs are observed at different locations, we assume, following the principle of Occam's razor, all three populations formed several Myr ago in one common gaseous disk rather than multiple disks at slightly different epochs \citep{BonnellRice2008, HobbsNayakshin2009, alig2013}. Their radial distance from Sgr A$^\star$ is at or outside the typical broad line region in AGN disks.  Moreover, 
AGN's spectroscopy indicates super-solar metallicity which signifies {\it ongoing} and 
{\it in situ} star formation and pollution in the proximity of SMBHs \citep{huang2023}. Theoretical models of AGN disks also suggest it may be a favorable site of star formation due to gravitational
instability \citep{Goodman2003, thompson2005} or star capture \citep{davies2020}.

%We simulate the dynamical evolution of these stars under the assumption they were formed \citep{Goodman2003, Nayakshin2007, levin2007} or captured and rejuvenated 
%\citep{syer1991, artymowicz1993, davies2020} several Myr ago in a single 
%(rather than multiple \citep{bonnell2008, hobbs2009, alig2013}) disk.
%We implant 500 stars with mass $M_\star=15 M_\odot$, negligible 
%initial $e_\star$ and an $a_\star$ distributed as $d {\rm N} / d a_\star \propto 
%a_\star^{-1.5}$ between inner ($a_{\rm in}=0.01$ pc) and outer ($a_{\rm out}=0.2$ pc) disk radii. This radial domain is comparable to those of the broad line region 
%in AGN disks where ongoing {\it in situ} star formation and pollution has been inferred from their super-solar metallicity \citep{huang2023}.  

We implant a total number $N_\star=500$ of coeval stars in one gaseous disk, on initially circular ($e_0 \sim 0$) Keplerian orbits, with an initial semi-major axis distributed $d {\rm N_{\star}} / d a_\star \propto a_\star ^{-0.5}$. We limit the initial semi-major axis distribution to $a_0 \geq a_{\rm in}=0.01$ pc $> > r_\bullet$ for marginally unstable disks \citep{Goodman2003} and the zone of avoidance \citep{burkert2024} (\S\ref{sec:complement}G). We also adopt the initial semi-major axis distribution that follows $a_0 \leq a_{\rm out}=0.2$ pc to avoid initial close encounters with the IMC, where $a_{\rm in}$ and $a_{\rm out}$ are the inner and outer boundaries of the star’s initial semi-major axis (Table \ref{tab:models}).
Between this set of inner ($a_{\rm in}$) and outer ($a_{\rm out}$) radii, the surface number density of the stars is 
assumed to be 
\begin{equation}
S_{\star} = {1 \over 2 \pi a_\star} {d {\rm N} \over d a_\star} 
= { N_\star \over 4 \pi a_{\rm out}^{0.5} a_{\star}^{1.5}} \frac{1}{[1-(a_{\rm in}/a_{\rm out})^{0.5}] }  ,
\label{eq:sstar}
\end{equation}
which is similar to the observed present-day stellar surface density $\propto a_\star^{-2}$
\citep{bartko2010} with a somewhat uncertain incompleteness factor. 
Additional simulations with $S_\star \propto a_{\star} ^{-2}$ (model {\it Steep}, Table \ref{tab:models}) 
%left Figure \ref{fig:IMF}) produces similar results as the {\it Fiducial} model (bottom right, Figure \ref{fig:e_peri_a0_t}). 
is also considered for a comparison.

We adopt an initial normal inclination ($i_\star ^\prime \equiv i_0 ^\prime=i_\star-i_0$) 
dispersion, extending from an aspect ratio $-H_{\star}/r_\star$ to $H_\star/r_\star$, with 
\begin{equation}
    \frac{H_\star}{r_\star} = \frac{H_0}{R_0} \left(\frac{r_\star}{R_0} \right)^{1/4},
\end{equation}
where $R_0$ is a {\it Fiducial} scaling radius, set to be $10^3$ AU and $H_\star(r_\star = R_0)=H_0=0.04 R_0$.
This prescription provides a small velocity dispersion for the stars (most of which reside
near the mid-plane). 
With this limited initial velocity dispersion, the stars'
Roche surface, with radius $r_{\rm R} = a_\star (M_\star/3 M_{\bullet}) ^{1/3}$, nearly
overlaps each other.
The Hill's radius of each star is comparable to or larger than the scaling height, $H$, and
their total mass is slightly less than the mass of the IMC. 
For the brief computational intervals (up to 9 Myr), we neglect the effects of mass loss due to stellar evolution. 
Their orbital eccentricity, $e_\star$, rapidly increases beyond $H_0/R_0 \sim 2.5 
r_{\rm R}/a_\star$ (\S\ref{sec:dynrelax}). The gravitational stability parameter for 
the star's disk \citep{toomre1964} follows
\begin{equation}
Q_\star \sim \frac{H_\star}{R_0} \frac{M_\bullet}{\pi a_0^2 S_\star M_\star} \sim {\mathcal{O} (10)} .
\end{equation}
The magnitude of $H_0$ and $H_\star$'s radial dependence does not influence our results.  

All stars are assigned a mass $m_\ast = 15~M_{\odot}$ to approximate the average observationally-inferred mass function for the S-stars, CWSs, and ODSs \citep{bartko2010, gillessen2017, habibi2017, 
vonfellenberg2022}. The total number $N_\star$ of stars and their total mass $N_\star M_\star 
(\leq M_{\rm IMC})$ are a factor of two larger than those of the known population. Here, we take into account the uncertain detection probability \citep{burkert2024} and approximate the effect of mass loss due to stellar evolution.  
One additional model with Kroupa %and Salpeter 
initial mass function (model {\it IMF} in the Table \ref{tab:models})
\citep{marks2012} for
the same $N_\star$ and total stellar mass 
%or b) a second population of 500 stars with 5$M_\odot$ have 
has been simulated.  
%They produce essentially the same results (right, Figure \ref{fig:IMF}) as the {\it Fiducial} model 
%(bottom right, Figure \ref{fig:e_peri_a0_t}).  

%With $N_\star=500$, stars' domain of influence ($a_\star \pm r_R$, where $r_{\rm R} 
%= a_\star (2 m_\star/3 M_{\rm SMBH})^{1/3}$ is the Roche radius of two neighboring stars) overlaps
%each other and their eccentricity rapidly increases beyond $H_0/R_0 \sim 2.5 r_{\rm R}/a_\star$
%(\S\ref{sec:dynrelax}). The gravitational stability parameter for the stars \citep{toomre1964}  
%$Q_\star \sim (H_0/R_0) M_\bullet/ \pi r^2 S_\star m_\star \sim {\mathcal{O} (10^2)}$.
%The magnitude of $H_0$ and the functional form of the radial dependence do not influence our results.  

%%%%%%%%%%%%%%%%%%%%%%%%%%%%%%%%%%%%%%%%%%%%%%%%
\subsubsection{Potential of the gaseous-disk}
\label{sec:gasdisk}

Our scenario is based on the conventional assumption that the S-stars, the CWSs, and the ODSs around them were formed in a thin gaseous disk within a fraction of a parsec from Sgr A$^\star$. Gravity from a depleting gaseous disk is also included. We adopt the assumption that outflow has transformed the gaseous disk into the Fermi bubble \citep{Su2010}. 

For computational convenience, we assume a power-law surface density distribution for the disk gas
\begin{equation}
    \Sigma_{\rm g} = \Sigma_0 \left(\frac{r}{R_0} \right)^{-3/2} e^{-t/\tau_{\rm dep}} ~\rm g/cm^2,
    \label{eq:sigma}
\end{equation}
where $\tau_{\rm dep}$ is the depletion timescale of the gas nebula, set as $2.5$ Myr, and the {\it Fiducial} surface density at $r = R_0 (=5 \times 10^{-3}$ pc) is chosen to be $\Sigma_0=800~ \rm{g/cm^2}$.  This prescription is chosen to be analogous to that prescribed for protostellar disks in most previous works \citep[e.g.,][]{nagasawa+2005, zheng+2017a, zheng+2020}, such that the sweeping secular resonance (SSR, \S \ref{sec:sweeping}) would pass through the clockwise disk stars (CWSs) region within several Myr.
It also has a radial dependence similar to that of AGN disks with marginal gravitational stability \citep{Goodman2003}. 

Condition for marginal gravitational (in)stability in the gaseous disk, 
$Q_{\rm g}= h_{\rm g} M_\bullet/ \pi \Sigma_{\rm g} r^2 \sim {\mathcal O} (1)$ \citep{safronov1960, toomre1964} is satisfied with the aspect ratio of the disk gas $h_{\rm g} \lesssim 3 \times 10^{-3}$ (which corresponds to a sound speed 1 km s$^{-1}$ at 0.1 pc). Such a thin disk structure is consistent with that inferred from the reverberation map for the AGN disc in NGC5548 \citep{starkey2023}, taking into account the height of the disk photosphere is several times its pressure scale height \citep{garaud2007}.

The total initial mass of the gaseous disk $M_{\rm disk} = 4 \pi \Sigma_0 R_0^2 (R_{\rm out}/R_0)^{1/2} \sim  10^4 M_\odot$ out to a disk radius ($R_{\rm out} \simeq 10^2 R_0 = 0.5$ pc $\geq (1+e_{\rm IMC})a_{\rm IMC}$) 
which contains all the young stars (including the S-stars, CWSs, and 
ODSs between $a_{\rm in}=0.01$ pc and $a_{\rm out}=0.2$ pc)
is slightly larger than the total mass ($N_\star M_\star = 7.5 \times 10^3 M_\odot$)  
of all the stars between $a_{\rm in}=0.01$ pc and $a_{\rm out}=0.2$ pc, and comparable to the mass of the IMC ($M_{\rm IMC}$).  
But, it is a small fraction of Sgr A$^\star$'s mass ($M_{\rm disk} < < M_\bullet$), 
and that is estimated for the Fermi bubble \citep{Su2010, zubovas2011}. 
More massive disks with larger $\Sigma_0$ can also be marginally
(un)stable with a larger aspect ratio $h_{\rm g}$.  
We also simulate a larger value of the gaseous surface density $\Sigma_0 = 8000~{\rm g/cm^2}$ in models {\it Massive Disk 0} and {\it Massive Disk 1}, considering $\tau_{\rm dep} = 2.5$ Myr and 1 Myr, respectively.  
%A larger value of $\Sigma_0$ would not affect the 
%strength of IMC's secular resonance, though the sweeping epoch would be somewhat delayed until $\Sigma_{\rm g}$ is reduced to sufficiently low values (item E, \S\ref{sec:complement}).  

Here we are mainly interested in the dynamical evolution at the Galactic center region ($r<R_{\rm out} = 0.5$ pc), where the potentials from the galactic disk and halo components can be ignored. We also exclude the contribution to the potential from the Galactic bulge because it is dominated by the self-gravity of the gaseous disk, especially in the interior ($<0.1$~pc) region and at the beginning of the disk depletion stage \citep{zheng+2020}. With a density scale height comparable to $H_\star$ in the 
direction normal to the disk plane, gravitational stability is well preserved. For this prescription ($\Sigma_{\rm g}$ in Equation \ref{eq:sigma}), the divergent gaseous disk potentials 
\citep{zheng+2020} for all stars are simplified to 
\begin{equation}
\Phi_{\rm \star, disk} \simeq  - 4\pi G \Sigma_{\rm g} r  .
\label{eq:f_star}
\end{equation}
 
We neglect the disk potential for IMC with an orbit that is inclined to the gaseous disk ($i_{\rm IMC} ^\prime = i_{\rm IMC} - i_{\star} \neq 0$), including the {\it Fiducial} model.  
But for an embedded IMC (with $i_{\rm IMC} ^\prime=0$), we adopt a modified disk potential 
%
%and
%
\begin{equation}
\begin{split}
 & \Phi_{\rm IMC, disk}   =  2 \pi G \Sigma_{\rm g} r \sum_{l=0}^{\infty}  \frac{A_l}{k}   \left[\left( \frac{r}{r_{+}}  \right)^{k} + \left( \frac{r_{-}}{r} \right)^{k}  \right] , \\
& {\rm with} \ \  A_l = \left[ \frac{(2l)!}{2^{2l} (l!)^2} \right]^2, \ \ \ \ {\rm and} \ \ \ \ k = \frac{4l+1}{2}.
\end{split} 
\label{eq:f_IMC}
\end{equation}
We assume a gas-depleted gap is induced by the IMC over the radial range between
$r_{\pm} = a_{\rm IMC} ( 1 \pm e_{\rm IMC}) (1 \pm (M_{\rm IMC}/M_{\rm SMBH})^{1/3} )$.

%%%%%%%%%%%%%%%%%%%%%%%%%%%%%%%%%%%%%%%%%%%
\subsection{Dominant Physical Processes}
\label{sec:dominant}

As an extension of our previous investigations, the dynamic interaction between young stars is fully incorporated. Under these conditions, modest eccentricity ($e_\star$) and warp of the CWSs are excited by stars' mutual gravitational interaction \citep{kocsis2011, kocsis2015}. It confirms that dynamical relaxation alone cannot simultaneously lead to the large observed eccentricity and inclination for most S stars and ODSs within their estimated age limit \citep{Freitag2006, Alexander2007, Madigan2011, rantala2024}. But, in the presence of IMC, its distant secular perturbation can introduce effective eccentricity and inclination excitation to disk of stars with initially circular orbits.
%Figure (\ref{fig:scenario}) provides a qualitative illustration of the dominant physical 
%mechanisms that regulate the evolution of stars near Sgr A$^\star$.

\begin{figure*}[htbp!]
\centering
\includegraphics[width=0.7\columnwidth]{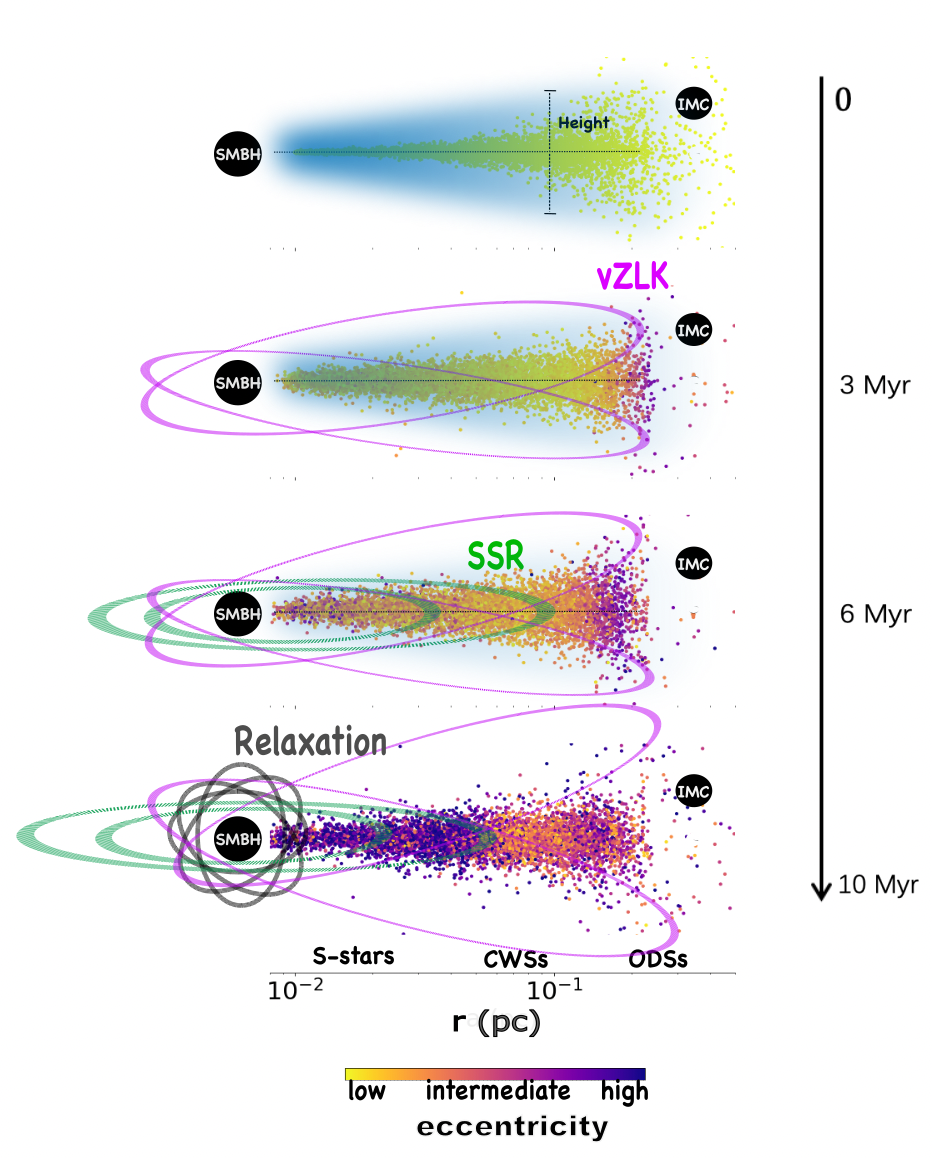}
\caption{A schematic illustration of the physical processes associated with the excitation of young stars 
in the proximity of Sgr A$^\star$ region. The central SMBH and an potential IMC are represented with black dots, while the blue 
shading indicates a depleting gaseous disk. Color dots show the eccentricity and inclination excitation 
of the S-stars, clockwise-disk stars (CWSs), and off-disk stars (ODSs). The yellow dots refer to 
low-eccentricity orbits, while the purple dots refer to high-eccentricity orbits. The top to bottom 
rows represent an evolutionary sequence over several Myr. Spatial domain and temporal epochs dominated 
by three different physical effects are labeled.}
\label{fig:scenario}
\end{figure*}

A schematic illustration of the physical processes associated with the excitation of young stars in the proximity of Sgr A$^{\star}$ in a unified model is shown in Figure \ref{fig:scenario} (For a detailed analysis, see discussions in \S\ref{sec:results}).
We briefly summarize three dominant physical effects that regulate the dynamical evolution of Galactic center stars.

%The IMC has a strong gravitational effect on its nearby disk stars. Its highly inclined orbit causing large modulations in $e_\star$ and $i_\star ^\prime$ (so-called von Zeipel-Lidov-Kozai, vZLK, effect). After few %($\sim 3$, see Figure \ref{fig:e_peri_a0_t}) 
%Myrs, IMC's vZLK rapidly excites $e_\star - i_\star ^\prime$ among the ODSs in its proximity and injects some into the S-stars domain. It also naturally generates an outer warp in the disk, as observed \citep{lockmann2009, bartko2009, 
%bartko2010, yelda2014}.
%Secular resonance occurs at the location where the disk and the IMC-induced precession frequencies match with the precession frequency of the IMC, leading to rapid eccentricity excitation for the young stars. 
%Subsequently, the so-called sweeping secular resonance (SSR) propagates as the location of secular resonance sweeps towards Sgr A$^{\star}$ over a vast region ($0.04-0.2$ pc) due to the depletion of the gaseous disk. The eccentricity of some resonant stars is excited towards unity with periapses reduced to a few Sgr A$^\star$'s gravitational radii \citep{zheng+2021}.

%%%%%%%%%%%%%%%%%%%%%%%%%%%%%%%%%%%%%%%%%%%%%%
\subsubsection{von Zeipel-Lidov-Kozai effect}
\label{sec:kozai}
In the dominant gravitational field of the Sgr A$^\star$, the secular perturbation of the IMC cumulatively modulates, over its successive orbital periods, nearby newly-formed stars in the disk to undergo large eccentricity-inclination oscillations. After a few Myr, this so-called von Zeipel-Lidov-Kozai (vZLK) effect \citep{vonzeipel1910,lidov1962,kozai1962,naoz2016, bhaskar2021}  excites large eccentricity-inclination for the ODSs and injects some of them into the S-stars domain close to Sgr A$^\star$. The IMC's perturbation also naturally generates an outer warp in the disk, as observed among the CWSs \citep{lockmann2009, bartko2009, bartko2010, yelda2014}. 

In our simulations, we consider a range of inclination $i_{\rm IMC}$ between the IMC orbital plane and the disk, including the {\it Fiducial} model with $i_{\rm IMC}^{\prime} = i_{\rm IMC} - i_{\star} =120^{\circ}$.
For an IMC with such a highly inclined orbit ($40^{\circ} \leq i_{\rm IMC} ^\prime \leq 140^{\circ}$), prolonged retention of the stars' longitude of periapse, $\varpi_\star$ (relative to that of the IMC), at some finite values \citep{innanen1997}. 
It can cumulatively lead to their large-amplitude of inclination and eccentricity 
%($i_\star^\prime -e_\star$) 
modulation through cumulative angular momentum
transport via the vZLK resonance on a characteristic time scale $\tau_{\rm vZLK}$. At the inner orbit of the IMC ($a \leq a_{\rm IMC}$), this characteristic timescale can be simplified as
\begin{equation}
\tau_{\rm vZLK} \simeq \frac{M_{\bullet} a_{\rm IMC} ^3}{M_{\rm IMC} a_\star ^3}  (1-e_{\rm IMC}^2) ^{3/2} P_\star \simeq 3 
\left( \frac{a_{\rm IMC}}{a_\star} \right)^{3/2} {\rm Myr} > > 
P_\star ,
\label{eq:taukozai}
\end{equation}
where $P_{\star}$ is the stars' orbital period. And the characteristic vZLK resonance timescale is much longer than the orbital period of the IMC ($\tau_{\rm vZLK} > > P_{\rm IMC} \sim 7.3 \times 10^3$ yr) \citep{valtonen2006}.
For those stars that are far away from the IMC with $a_0 \leq 10~a_{\rm in}$, the eccentricity excitation via the vZLK effect  is limited during massive stars' lifespan ($\tau_\star \sim 10^7$yr) as their characteristic vZLK resonance timescale is even longer than their lifetime ($\tau_\star 
\leq \tau_{\rm vZLK}$). But, for the O-B stars in the proximity of the IMC (with $a_\star \geq 0.1$ pc, Figure \ref{fig:scenario}), the magnitude of the characteristic vZLK resonance timescale ($\tau_{\rm vZLK}$) is a few Myr, comparable to their lifespan.  Provided the initial inclination is sufficiently large (with sin$^2 i_{\rm IMC}^{\prime} \geq 2/5$ or
$40^{\circ} \lesssim i_{\rm IMC}^{\prime} \lesssim 140^{\circ}$), the eccentricity of some disk stars can also be excited from negligible values to unity ($e_\star \rightarrow {\mathcal O} (1)$ ) by the vZLK resonances \citep{naoz2016, bhaskar2021} over 
$\tau_{\rm vZLK} \leq \tau_\star$ \citep{innanen1997},
provided that the perturbation from other stars is negligible (\S\ref{sec:dynrelax}). 

In the absence of IMC's precession, 
the relative longitude of periapse ($\varpi_\star$) clusters around some certain values as 
sin$^2 (\varpi_\star) \sim 2/5~{\rm sin}^2 i_{\rm IMC}$ or $\varpi_\star \sim \pm \pi/4$ 
(or $\pi \pm \pi/4$), without a preferred longitude of ascending node ($\Omega_\ast$),
during the eccentricity excitation process \citep{innanen1997}. These phase-space ``attractors'' are likely to be
slightly modified by IMC's precession.  
Due to the large characteristic vZLK timescale $\tau_{\rm vZLK}$, this mechanism is ineffective for the young S-stars, which are more
distant to the IMC. 

%%%%%%%%%%%%%%%%%%%%%%%%%%%%%%%%%%%%%%%%%%
\subsubsection{Sweeping secular resonance}
\label{sec:sweeping}
%Moreover, the IMC's secular resonance occurs at the location where the precession frequencies of the stars (mainly induced by the modified gaseous disk gravity and the IMC, and the secular interaction timescale of IMC can be simplified as $\tau_{\rm SI} \simeq (a_{\rm IMC}/a_\star)^{1/2} P_{\rm IMC} M_\bullet/M_{\rm IMC}$) match with the precession frequency of the IMC \citep{ward1981, nagasawa+2005, zheng+2020}. During the gas disk depletion, the IMC's so-called sweeping secular resonance (SSR) propagates as the location of secular resonance sweeps towards Sgr A$^{\star}$ over a vast region ($0.04-0.2$ pc). Stars along the path of the sweeping secular resonance of the IMC undergo rapid eccentricity excitation towards unity as their orbital perigee approaches the central SMBH ($r_p \rightarrow r_\bullet$) \citep{zheng+2021}.

Although the gas-disk potential does not significantly modify the Keplerian velocities, it does lead to the precession of stars' longitude of periapsis ($\varpi_\star$) and ascending node ($\Omega_\star$) as well as those of a co-planar eccentric IMC ($\varpi_{\rm IMC}$ and $\Omega_{\rm IMC}$). 
In the {\it Fiducial} model, although we neglect the contribution of gas-disk potential to the dynamics of the (inclined) IMC due to its large relative inclination to the gaseous disk ($i_{\rm IMC}^{\prime} \equiv i_{\rm IMC} - i_{\star}$), the mutual secular interaction between young stars and the IMC also induces them to precess.  
Secular resonance occurs at the location where the gaseous disk and the IMC-induced precession frequencies (on secular interaction timescale $\tau_{\rm SI} \simeq (a_{\rm IMC}/a_\star)^{1/2} P_{\rm IMC} M_\bullet/M_{\rm IMC}$) \citep{zheng+2021} match 
with the precession frequency of the IMC \citep{ward1981, nagasawa+2005, zheng+2020}.  
%Here, we examine how this secular resonant interaction leads to eccentricity excitation for the young stars \citep{zheng+2021}. 

Since the gas-disk potential is determined by its surface density distribution $\Sigma_{\rm g} (r, t)$ (Equation \ref{eq:f_star}), the location of IMC's secular resonance is also a function of surface density $\Sigma_{\rm g}$ at any given time. 
As stars' precession rate decreases
with the depletion of the disk gas, the location of secular resonance sweep towards Sgr A$^\star$, the IMC's so-called sweeping secular resonance (SSR) propagates
over a vast region ($0.04-0.2$ pc). The eccentricity of these resonant stars is excited towards unity ($e_\star \rightarrow 1$) with periapses reduced to a few Sgr A$^\star$'s gravitational radii ($r_p \rightarrow r_\bullet$), while their semi-major axes are preserved \citep{zheng+2021}.

%%%%%%%%%%%%%%%%%%%%%%%%%%%%%%%%%%%%
\subsubsection{Dynamical relaxation}
\label{sec:dynrelax}

Both the vZLK effect and SSR of the IMC do not directly lead to significant changes in $a_\star$.
Moreover, the SSR does not reach the inner region ($\lesssim 0.1$ pc), and the stars formed there do not directly attain nearly parabolic orbits. 
Nevertheless, eccentricity of some stars with large semi-major axis, e.g., $a_\star \geq 0.1$ pc, may be excited to sufficiently large values for them to venture into the S-stars' proximity 
during their perigee passage. 
The conventional two-body relaxation timescale $\tau_{\rm rlx}$ for nuclear-cluster stars with nearly isotropic velocity dispersion
\citep{binney1987} is generally much longer than the age of the S-stars and CWSs \citep{yu2007, kocsis2015}, although the vZLK effect of the IMC may significantly enhance this process \citep{naoz2022}.

A disk of newly-formed, co-orbiting stars with nearly circular orbits undergo stellar-disk relaxation due to their mutual gravitational interaction, analogous to planetesimals in protostellar disks \citep{kokubo1998, ida2004}.  
%At $a_\star$, 
The velocity dispersion ($\sigma_\star$) \citep{palmer1993, aarseth1993} of stars increases at a rate 
${\dot \sigma}_\star ^2 \equiv d \sigma_\star ^2 / d t \simeq S_\star \Omega_k^5 r_{\rm R}^6 / \sigma_\star ^2$,
where $\Omega_k = (G M_{\rm SMBH} / a_\star ^3) ^{1/2}$ is the Keplerian angular frequency 
and $r_{\rm R}$ is the Roche radius of two neighboring stars.
Substituting an average $ \langle e_\star^2 \rangle \simeq \sigma_\star^2 a_\star / G M_{\bullet}$, the eccentricity of stars  at $a_\star$ is excited on a time scale
\begin{equation}
\begin{split}
\tau_{\rm e} & \simeq {\sigma_\star^2 \over {\dot \sigma}_\star^2 } = {9 M_{\bullet}^2 \over 2 m_\star^2 }
{\langle e_\star ^2\rangle ^2  \over N_\star} \left( 1 - {a_{\rm in} ^{1/2} \over a_{\rm out} ^{1/2}} \right) \left(\frac{a_{\star}}{a_{\rm in}} \right)^{1/2} P_{\star} \\
& \sim 20 
\langle e_\star ^2 \rangle ^2 \left( {a_\star \over a_{\rm in}} \right)^2 {\rm Gyr} . 
\label{eq:taue}
\end{split}
\end{equation}

The stellar-disk relaxation timescale is much shorter than the conventional two-body relaxation timescale ($\tau_{\rm e} < < \tau_{\rm rlx}$) for those newly formed stars with small ${\langle e^2_\star \rangle}$. 
Due to the relatively high stellar density and short orbital periods, stellar-disk relaxation alone can induce indigenous stars with an initial semi-major axis around the inner boundary ($a_0 \sim a_{\rm in}$) to diffuse their semi-major axis ($a_\star$) and attain an average eccentricity distribution $\sqrt {\langle e^2_\star \rangle} \sim 0.1-0.2$ within their lifespan ($\tau_\star \leq 10$ Myr).
%(top, left Figure \ref{fig:rp_timescale}).   
As the average eccentricity ($\sqrt {\langle e^2_\star \rangle}$) increases, the small-eccentricity approximation
for the stellar-disk relaxation timescale becomes inadequate. 
Neither stellar-disk relaxation nor conventional two-body relaxation can explain the high eccentricities ($e_\star \sim 0.5-1$) of the S-stars and ODSs, as both timescales exceed the relevant age constraints ($\tau_{\rm rlx} > \tau_{\rm e} > \tau_\star$) \citep{binney1987, yu2007, kocsis2015}.

Nevertheless, the vZLK effect and the SSR mechanism operate on viable timescale ($\tau_{\rm vZLK}, \tau_{\rm SI} < \tau_\star$) for orbits with semi-major axes $a_0 \geq 0.1$ pc, providing a pathway to excite high stellar eccentricities ($e_\star$). 

Under the dominant gravity of Sgr A$^\star$, the S-stars, CWSs, and ODSs on eccentric orbits undergo resonant relaxation (RR) process \citep{rauch1996, yu2007, hopman2006, kocsis2011, kocsis2015}. The characteristic timescale for changes in angular momentum magnitude is $\tau_{\rm RR, e} \sim 3 P_{\star} M_\bullet/M_\star$ \citep{murray2000}, while vector resonant relaxation drives reorientation on a timescale 
$\tau_{\rm RR, \varpi} \sim \tau_{\rm RR, e}/5 N_\star^{1/2} (<r)$
at a distance $r$ \citep{yu2007, kocsis2015}.
For the innermost S-stars (with $a_\star \lesssim 10^{-2}$ pc), 
the magnitude and orientation of the angular momentum vector  evolve on a resonant-relaxation timescale 
$\tau_{\rm RR} \sim 10^{6-7}$ yrs. This process may lead their orbital distribution toward an isotropic velocity dispersion.

For the more distant CWSs and ODSs ($a_\star \sim 0.04-0.2$ pc), the timescales for both stellar-disk relaxation and resonant relaxation exceed their stellar age ($\tau_{\rm RR, e} \geq \tau_\star$).
%and resonant relaxation alone cannot 
%excite ODSs' high-$e_\star$ \citep{Alexander2007}.  
%Nevertheless, 
Therefore, the relatively small eccentricity and inclination, including the warp of the CWSs, may be attributed to resonant relaxation \citep{kocsis2011, kocsis2015, yelda2014}. But, there may not be adequate time for the young ODSs to attain their relatively large eccentricities and inclinations through resonant relaxation alone \citep{Alexander2007}.  

Moreover, resonant relaxation induces cumulative angular momentum (but not energy) 
exchange between stars.  In the absence of alterations to the semi-major axis ($a_\star$), this modulates orbital eccentricity and inclination ($e_\star$ and $i_\star$), similar to that
excited by their mutual vZLK interaction \citep{naoz2022}.
For S-stars with sufficiently large eccentricity, the vector resonant relaxation timescale ($\tau_{\rm RR, \varpi}$) is shorter than their age ($\tau_\star$), allowing 
%(top right Figure \ref{fig:rp_timescale})
their orbital planes to become isotropically distributed. But,  
the transition from an initial state of circular, co-planar
orbits to their observed eccentricity-inclination distribution 
requires an adequate angular momentum deficit (AMD) 
\citep{laskar1997} carried nearby highly-eccentric
intruding stars (Figure \ref{fig:scenario}). This
required AMD cannot be generated within star's age ($\tau_\star$), 
merely by star-disk or two-body relaxation, but it 
is attainable through IMC's vZLK perturbation and 
sweeping secular resonance (\S\ref{sec:mechanisms}) as the
IMC can provide an infusion of angular momentum deficit.

%These estimates again suggest diverse dynamical evolution for the three populations
%of stars around \sgra.

%\subsubsection{Synergy and Competitiveness}
%\label{sec:synergy}
  
%(bottom panels Figure \ref{fig:rp_timescale}).

%%%%%%%%%%%%%%%%%%%%%%%%%%%%%%%%%%%%%%%%%%%%%%%%%%%%%%%%%%%%%%%%%%%%%%%%%%%%%%%%%%%%%%%%%%%%
\subsubsection{Synergy between IMC's perturbation and relaxation}
\label{sec:synergy}

%Figure \ref{fig:scenario} qualitatively illustrates the diverse evolutionary pathways of young stars around the SMBH Sgr A$^\star$. 

In the proximity of the IMC, the combined effects of the vZLK and SSR cumulatively excite stellar eccentricity to high values,  $e_\star \rightarrow {\mathcal O} (1)$.  Some stars attain peri-center distances $r_{\rm p} < a_{\rm in}$, and carry with them an angular momentum deficit (AMD) infusion during their peri-center passages through the inner S-star domain, where the vector resonant relaxation timescale ($\tau_{\rm RR, \varpi}$) becomes shorter than the stellar lifetime ($\tau_\star$).  The magnitude and orientation of stellar angular momentum vector (orbit-normal) evolve under resonant relaxation, leading to an isotropic velocity dispersion ($\sigma_\star$) for the S-stars. But, the hierarchy of timescales governs the overall dynamics: for those stars with semi-major axis $a_\star \leq (M_\star/M_{\rm IMC})^{1/3} a_{\rm IMC} \sim 0.04$ pc, the vector resonant relaxation is dominant rather than the vZLK effect as $\tau_{\rm RR, \varpi} \leq \tau_{\rm vZLK}$.
And the vector resonant relaxation timescale is even shorter than the stellar lifespan ($\tau_{\rm RR, \varpi} \leq \tau_{\star}$) for those stars that satisfy $a_\star \leq 
(M_\star/M_{\rm IMC})^{1/2} a_{\rm IMC} \sim 0.014$ pc. Consequently, within the transitional region near the CWSs domain, resonant relaxation can interrupt the secular evolution, thereby 
limiting the orbital excitation ($e_\star-i_\star$) for some stars
(see \S\ref{sec:mechanisms} \& \S\ref{sec:orientation}).

\section{Analysis of the computational models}
\label{sec:results}

\begin{figure*}[ht!]
\centering
\includegraphics[width=0.6\columnwidth]{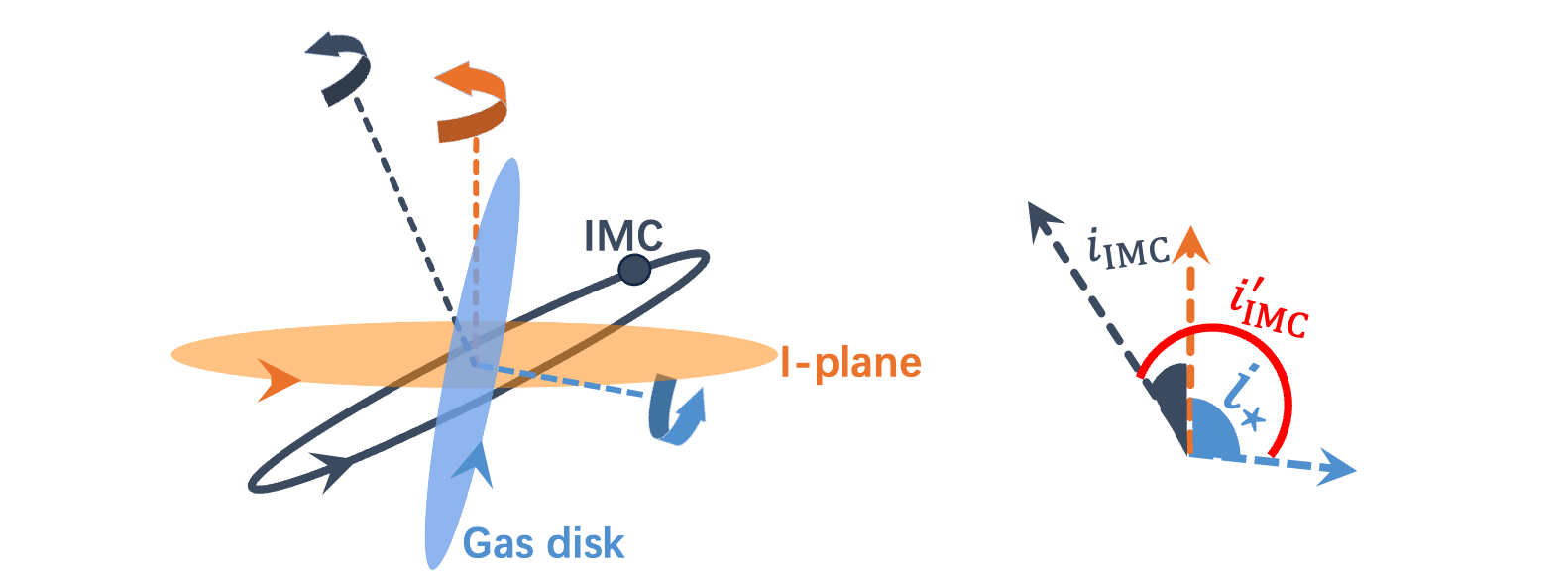}
\includegraphics[width=0.45\columnwidth]{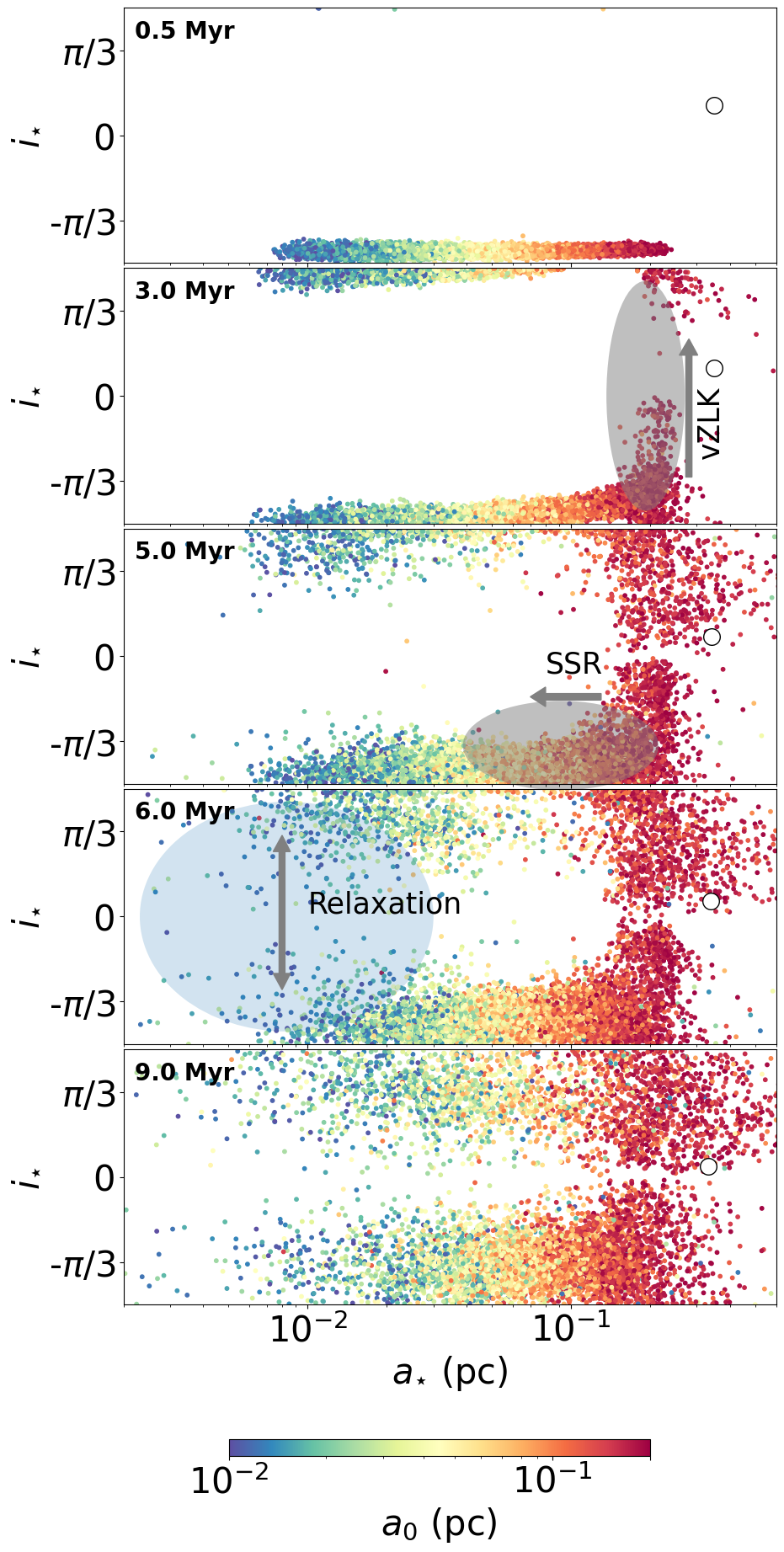}
\includegraphics[width=0.45\columnwidth]{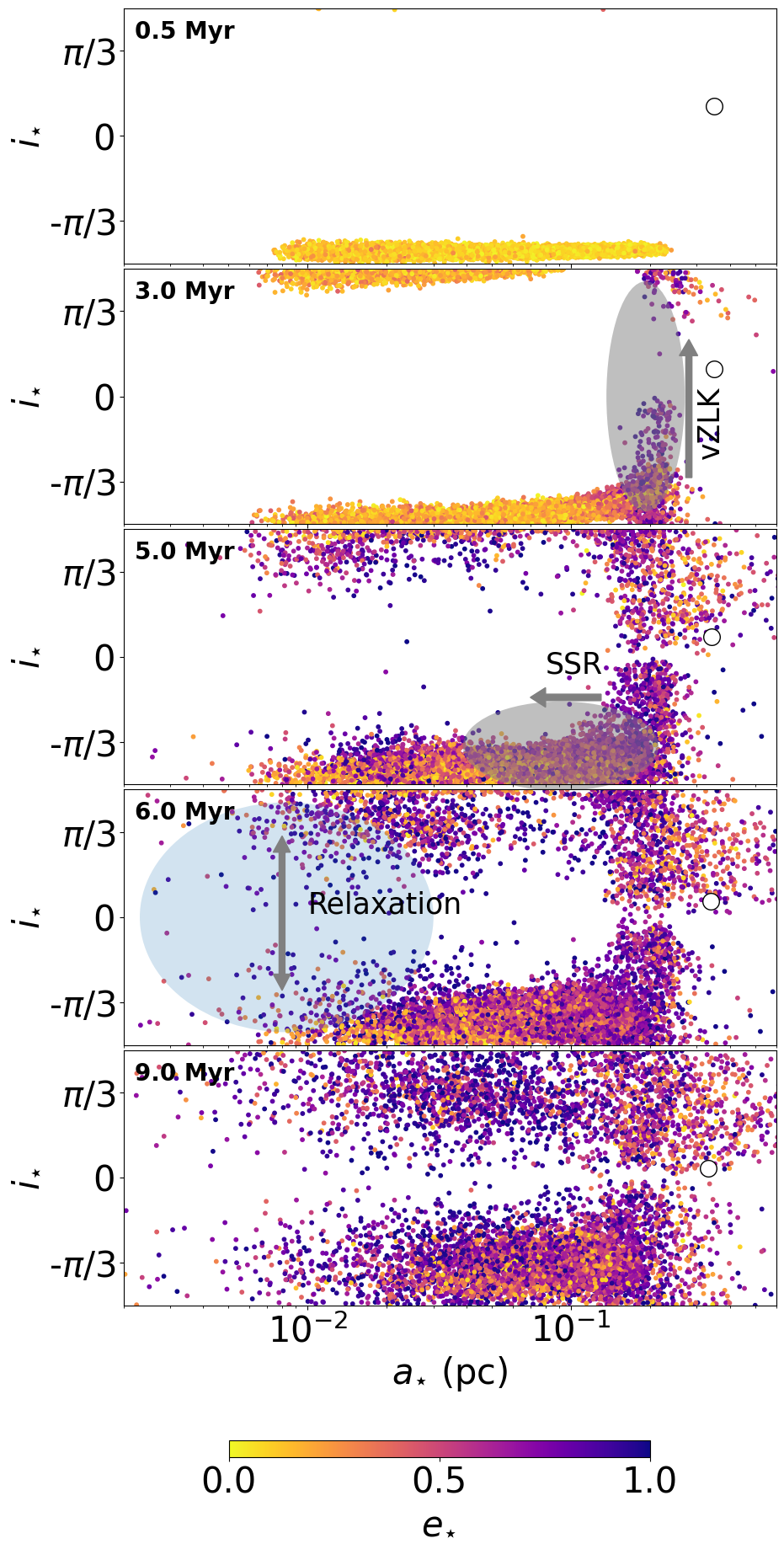}
\caption{The gas-disk, IMC-orbit, and total initial-AM planes (top-left) are shown with relative inclination angles (top-right).
Evolution of stars' $i_\star-a_\star$ distribution
with different $a_0$ (left) and current $e_\star$ 
(right). vZLK-, SSR-, and relaxation-dominant regions 
are highlighted with light-grey (vertical $3$ Myr, horizontal $5$ Myr) and blue ($6$ Myr) ellipses respectively.
}
\label{fig:a_inc_t}
\end{figure*}

\begin{figure*}[ht!]
\centering
\includegraphics[width=1\columnwidth]{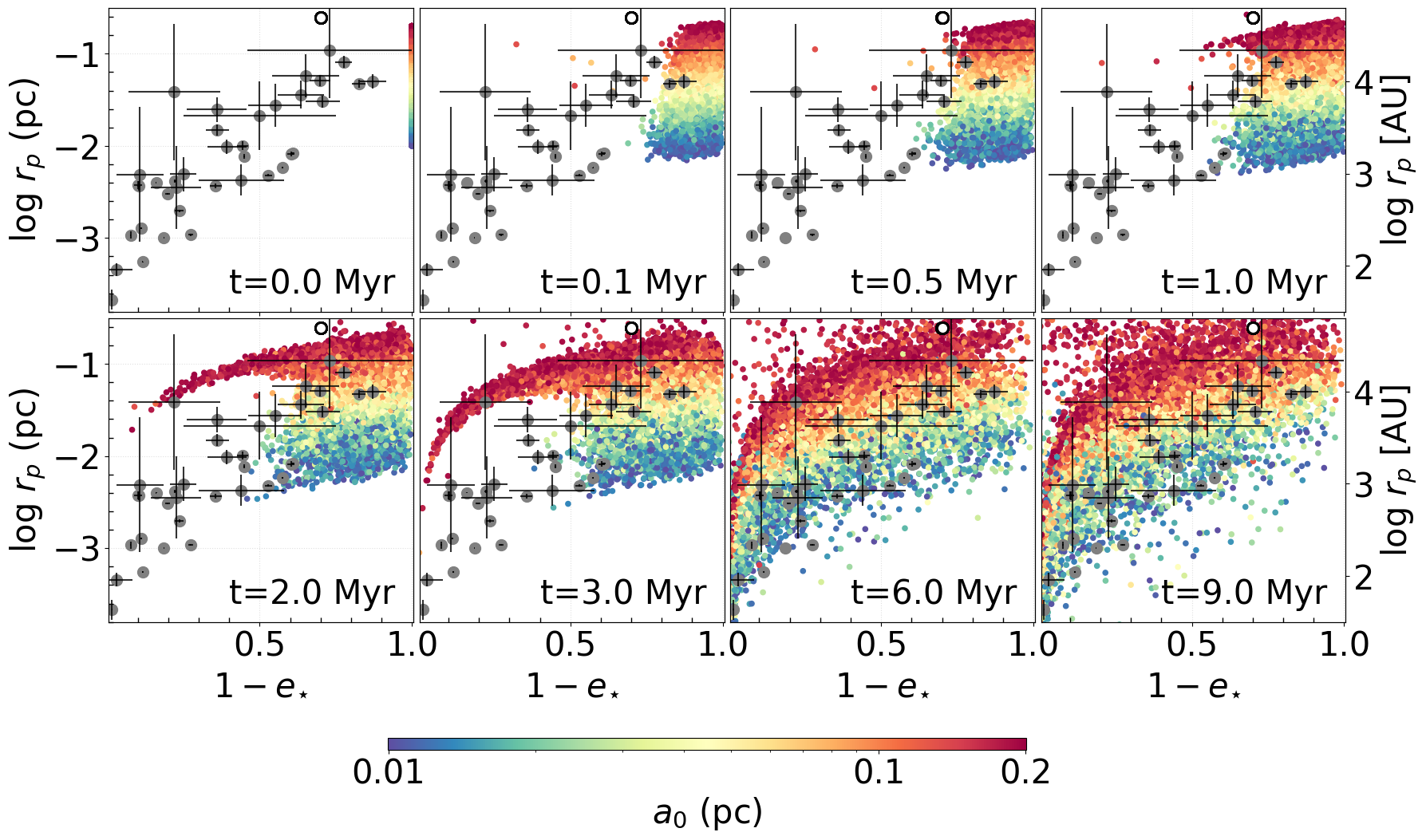}
\caption{The simulated results of the {\it Fiducial} model, which highlight the divergent evolution of these young disk stars 
around Sgr A$^\star$. The filled dots depict the perigee and eccentricity-related orbital parameters ($1-e$) at various epochs ($t = 0, 
0.1, 0.5, 1, 2, 3, 6, 9$ Myr). The black-open circle shows the orbits of an IMC. The color bars label the initial location ($a_0$) of these 
disk stars. Grey dots with error bars indicate the observed distribution of some S-stars at $\sim 0.003-0.18$ pc \citep{burkert2024}.}
\label{fig:e_peri_a0_t}
\end{figure*}

\begin{figure*}[ht!]
\centering
\includegraphics[width=1.\columnwidth]{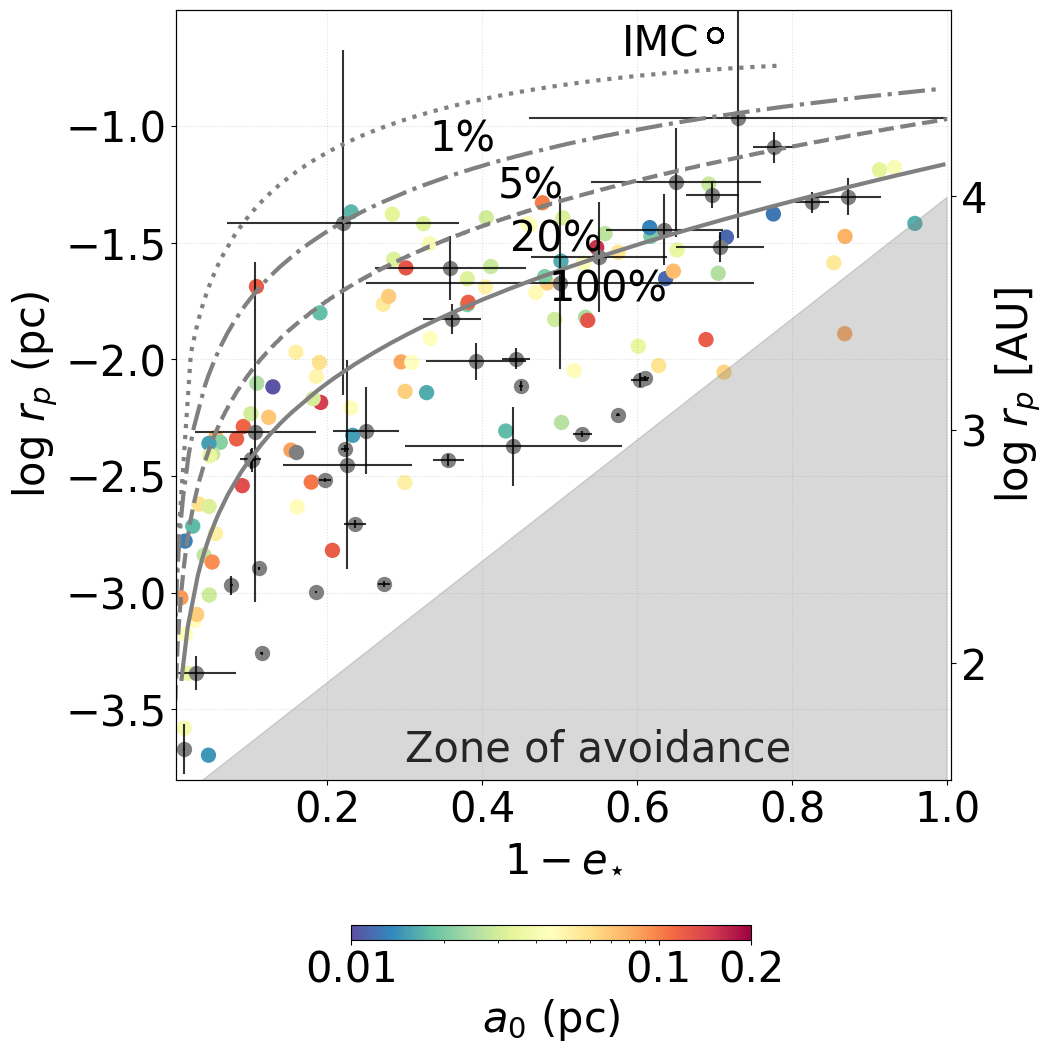}
\caption{ The simulated results of the {\it Fiducial} model, in which the IMC is $120^{\circ}$  inclined to the mid-plane. The filled dots depict the Peri-center distance $r_p$ and eccentricity-related orbital parameters ($1-e$) of selected stars at the final epoch ($t = 9$ Myr) after 
taking into account of their detection probability \citep{burkert2024}. Color bars label their initial 
location ($a_0$). The S-stars observed values (black dots 
with error bars) are shown against a zone
of avoidance (light-grey patch) \citep{burkert2024}. The grey solid, dashed, dashdot and dotted lines refer to $100\%$, $20\%$, $5\%$, and $1\%$ detection probability, respectively. 
The simulation reproduces the observed distribution.}
%The dashed black line marks the boundary between detected S-stars and a zone of avoidance shown in a companion paper \citep{Burkert+2023}.}
 \label{fig:e_peri_a0_t9}
\end{figure*}

\begin{figure}[ht!]
\centering
\includegraphics[width=0.7\columnwidth]{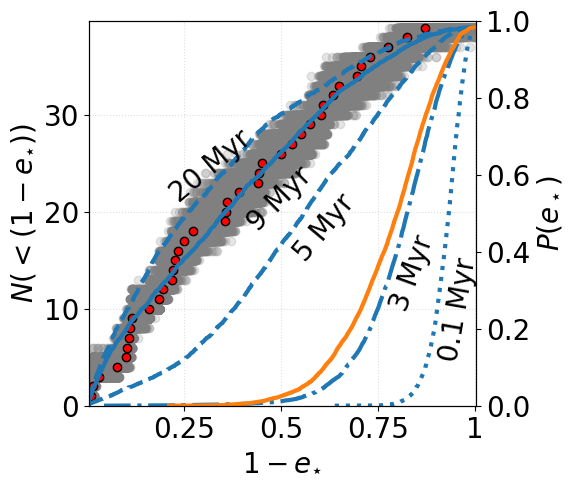}
\caption{The cumulative and normalized probability eccentricity distribution $P (e_\star)$ of 
observed S-stars (red dots with grey error-bars 
for $1 \sigma$ mean square variation \citep{burkert2024})
and
all the simulated stars (weighted by
detection probability, Figure \ref{fig:e_peri_a0_t9}) at various epochs ($t=0.1$ (dotted), $3$ (dot-dashed), $5$ (dashed), $9$ (solid)  and $20$ (dashed) Myr). 
The yellow line shows the {\it No IMC} model at $9$ Myr for comparison. Its mismatch 
with the red dots indicates that
stellar relaxation alone cannot reproduce the observed $e_\star$ distribution.}
\label{fig:e_cul_t}
\end{figure}

\begin{figure*}[ht!]
%\centering
\includegraphics[width=1.0\columnwidth]{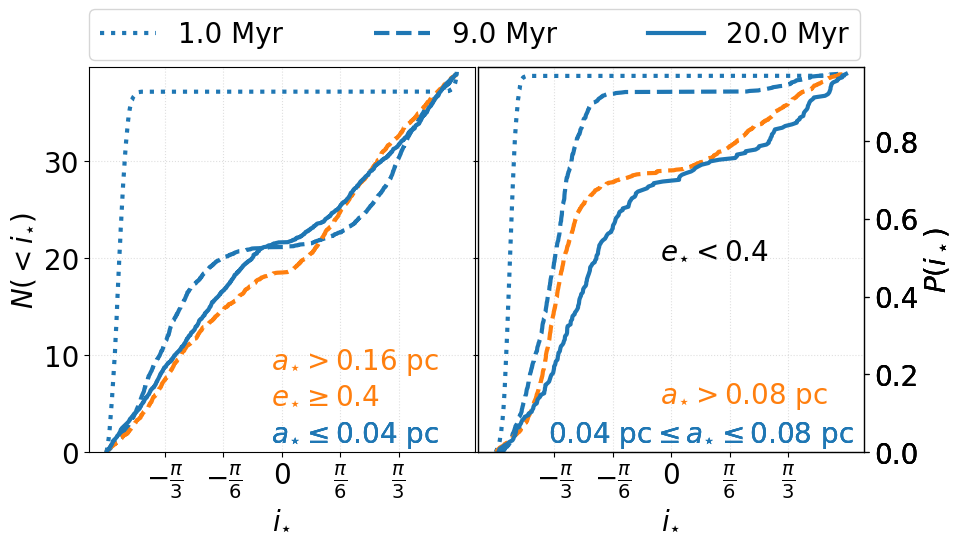}
\caption{Evolution of $P(i_\star)$ in 1) the S-star $(a_\star \leq 0.04$ pc) domain 
(blue, left panel); 2) inner $(0.04$ pc $\leq a_\star \leq 0.08$ pc, blue) and 
outer ($a \geq 0.08$ pc, red) CWS (with $e_\star \leq 0.4$) domains (right panel);
and 3) outer ODS ($a_\star \geq 0.16$ pc and $e_\star \geq 0.4$) domain (red, left panel).
At 1 Myr (dotted lines), the natal disk structure is intact in all regions.  S-stars'
$i_\star$ distribution is mostly isotropic after 9 Myr (dashed line) and totally 
uniform (with a diagonal line) after 20 Myr (solid line). Also at 9 Myr, the inner 
CWSs retain some initial disk signatures and the outer CWSs show signs of a warp 
with a gradual $i_\star$ distribution.  In comparison, the outer ODSs' dissimilar
$i_\star$ distribution alludes to a pseudo disk, cf Figure \ref{fig:cws_ods}. 
}
\label{fig:i_cul_t_a}
\end{figure*}

\begin{figure*}[ht!]
    %\centering
    \includegraphics[width=0.5\columnwidth]{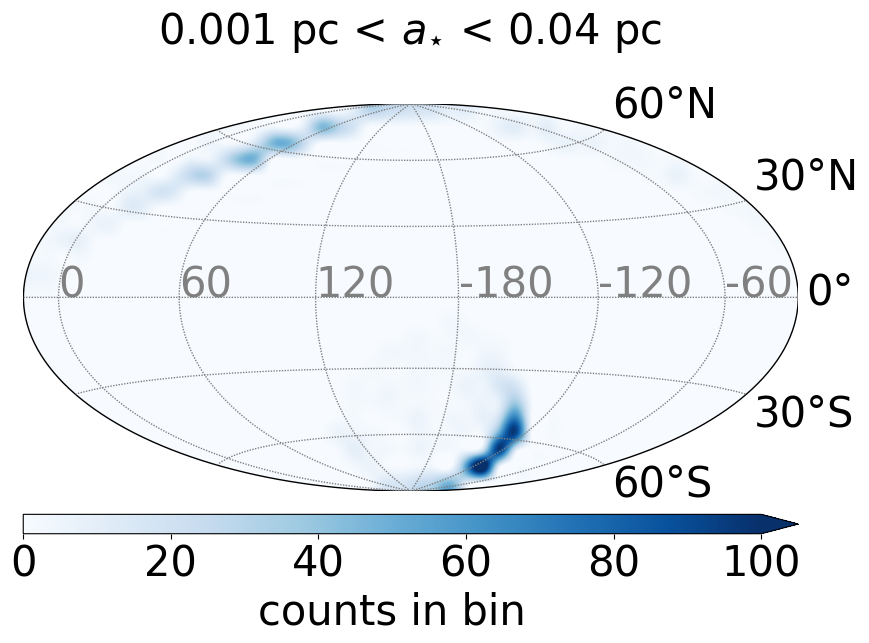}
    \includegraphics[width=0.5\columnwidth]{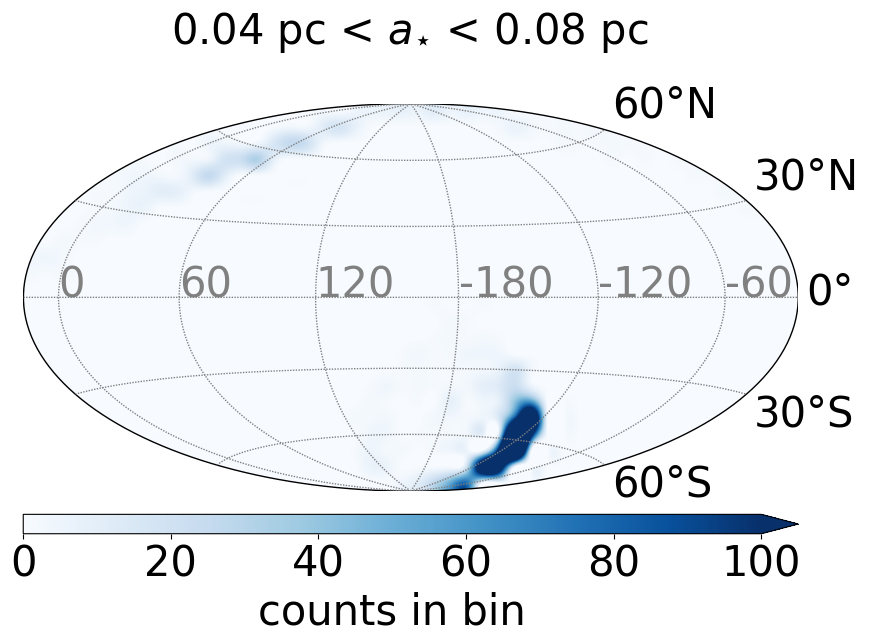}
    \includegraphics[width=0.5\columnwidth]{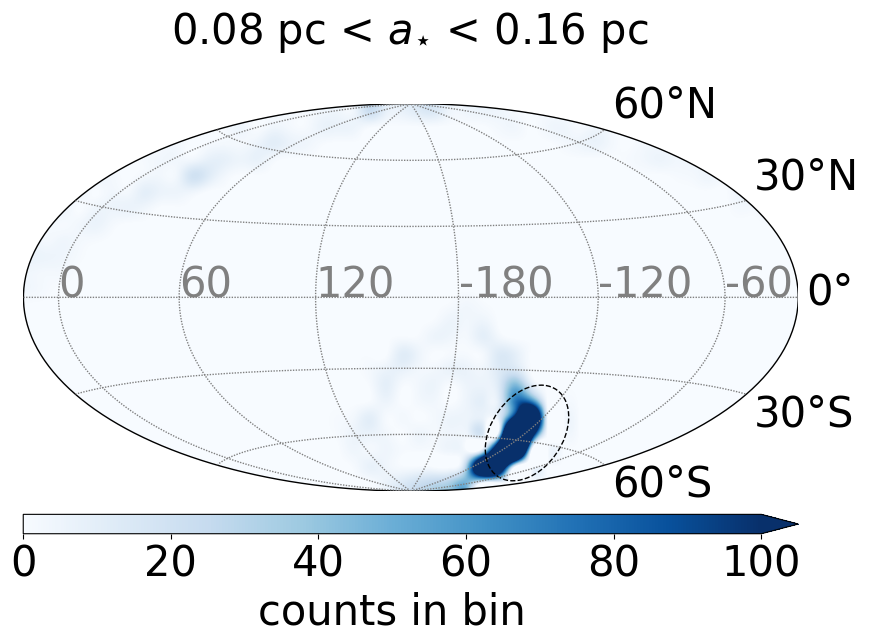}
    \includegraphics[width=0.5\columnwidth]{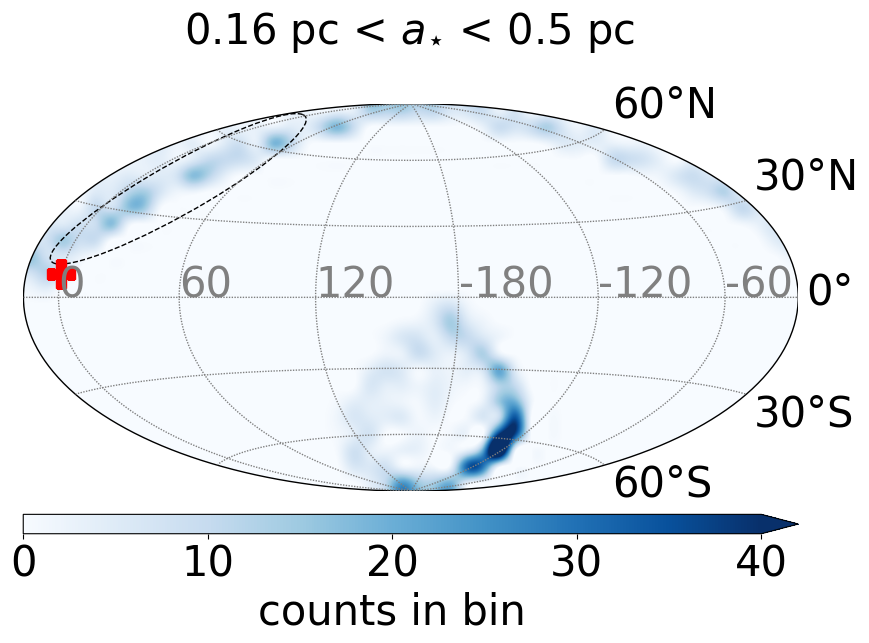}
    \caption{The ($i_{\star}$, $\Omega_{\star} - \Omega_{\rm IMC}$) distribution for S-stars ($a_\star \leq 0.04$ pc upper left),
    CWSs' inner (0.04 pc $\leq a_\star \leq 0.08$ pc upper right) and outer (0.08 pc $\leq a_\star \leq 0.16$ pc, 
    lower left) components, and the ODSs (0.16 pc $\leq a_\star \leq 0.5$ pc lower right).  
    Stars' $i_\star$ is relative to the initial invariant plane (total initial-AM plane). The color intensity corresponds
    to stars' phase-space concentration. The inner CWSs are much more concentrated in
    a patch than the S-stars.  The more extended $i_\star-\Omega_\star$ distribution 
    indicates a warped-disk structure (highlighted by a 
    dotted ellipse in the lower left panel).  The patchy distribution (highlighted by the dotted ellipse in
    the lower right panel)
    of the ODSs suggests a pseudo disk which is highly inclined from the disk of the CWSs but extending to 
    that of he IMC's orbit which is shown with a red cross.}
    \label{fig:cws_ods}
\end{figure*}

\begin{figure*}
\centering
\includegraphics[width=1\columnwidth]{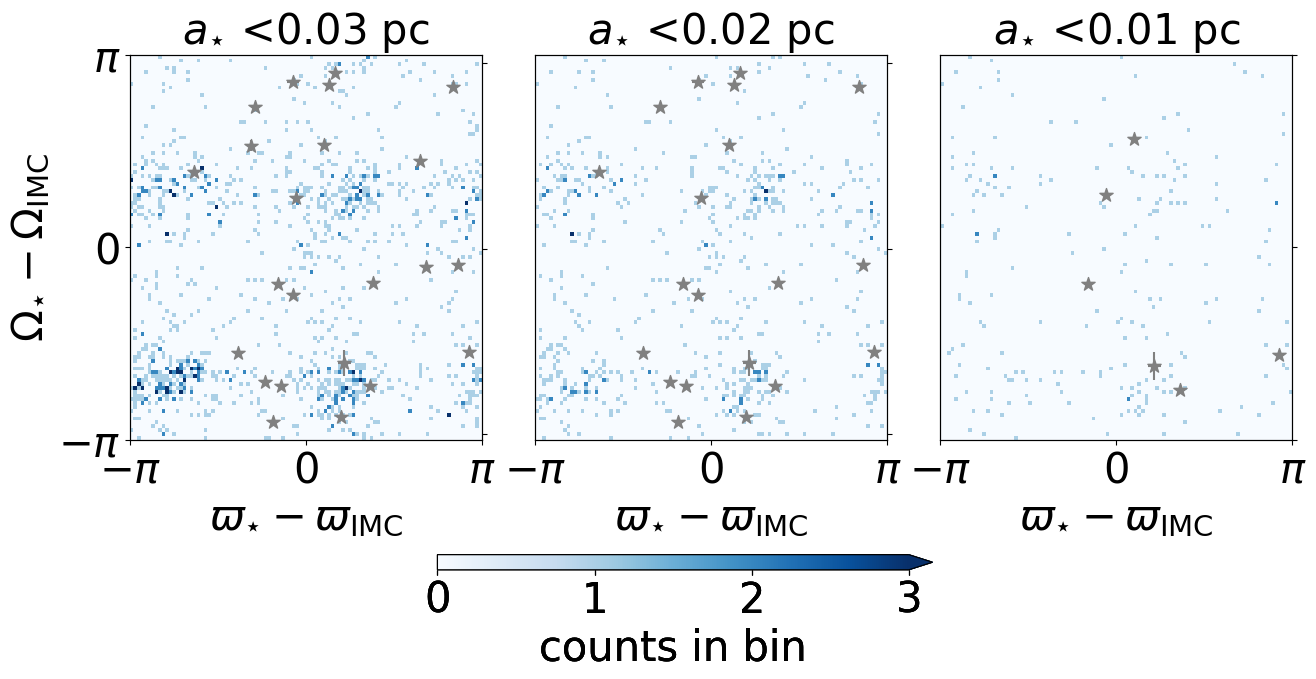}
\includegraphics[width=0.75\columnwidth]{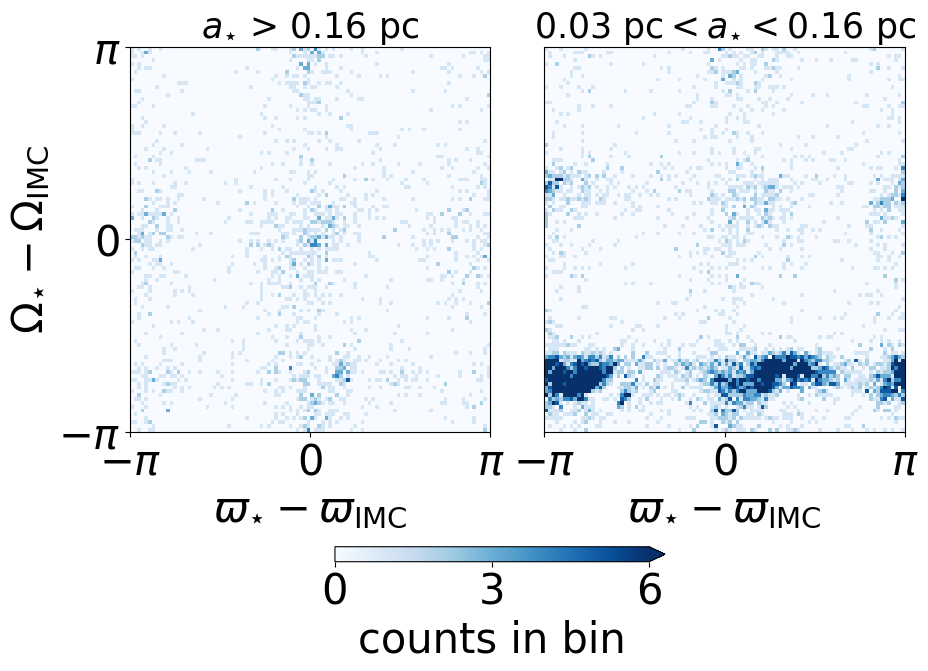}
\caption{The ($\varpi_{\star} - \varpi_{\rm IMC}$, $\Omega_{\star} - \Omega_{\rm IMC}$) distribution of simulated stars after 9 Myr evolution for each component, where $i_\star$ is relative to the initial invariant plane (total initial-AM plane), and $\Omega_{\star}$ and $\varpi_{\star}$ are longitude of ascending node and periapse of simulated stars. The $\Omega_{\rm IMC}$ and $\varpi_{\rm IMC}$ are relative to the initial reference plane. 
The top panel compares the final orbital parameters of simulated stars within $0.04$ pc, $0.02$ pc, and $0.01$ pc, respectively. The observed stars in the same region are represented with white star symbols. 
The bottom panels refer to the highly excited ODSs (left) and CWSs (right) populations, which may be related to the vZLK resonance and SSR, respectively. 
} 
\label{fig:pomegaomega0}
\end{figure*}

In this section, we provide detailed numerical simulations to verify our dynamical
scenario.  The combined contributions of the different physical processes manifested 
in the numerical simulations, unfolds as follows.

\subsection {The fiducial model}
\label{sec:fiducial}

A {\it fiducial} model is presented as a case study.  The parameters for 
the {\it fiducial} model (Figures \ref{fig:a_inc_t} $-$ \ref{fig:pomegaomega0}
) are chosen through an extensive and systematic parameter analysis with many sets of 
numerical simulations. Through this study, we dissect contribution from
different physical processes. For each model, we use $\sim 20$ independent sets of 
initial seeds to boost the statistical significance 
(\S\ref{sec:mechanisms}-\ref{sec:complement}).
and cast constraints on the mutual inclination between the IMC and young disk stars ($i_{\rm IMC}^\prime = i_{\rm IMC} - i_{\star}$, see also in \S\ref{sec:imcinclination}).

In the {\it Fiducial} model, which have an initial semi-major axis $a_0$ distribution between $a_{\rm in}=0.01$ and $a_{\rm out}=0.2$ pc (\S\ref{sec:diskstars}), all young stars are embedded in the same gas disk, inclined to the IMC's orbit by 120$^{\circ}$. This system of disk stars evolves under the influence of their mutual gravitational interaction as well as the IMC's secular perturbation and the gas disk's gravity.  

The colors of the stars in Figures \ref{fig:scenario} and the right panel of \ref{fig:a_inc_t} show the star's orbital eccentricity.
Between $1-3$ Myr, the eccentricity of a fraction of disk stars (with $a_0 \sim 0.1-0.2$ pc) is excited towards unity by the IMC's vZLK (\S\ref{sec:kozai}) and SSR (\S\ref{sec:sweeping}) mechanisms.
%(\S\ref{sec:kozai}) and SSR (\S\ref{sec:sweeping}).  
The relative contribution of these resonances is further characterized separately by the stars' inclination ($i_\star$) and longitude of periapsis ($\varpi_\star$).  
Due to the depletion of the disk gas, the inward propagation of IMC's secular resonance over a wide region is illustrated by the distinctive eccentricity difference between the purple (with $a_0 \sim 0.2$ pc) and the orange (with $a_0 \sim 0.1$ pc) dots. 
As shown in our previous detailed analysis \citep{zheng+2020,zheng+2021}, the continuous increase of orbital eccentricity may lead to eccentricities $e_\star \approx 0.3$ after $6$ Myr. This process provides a promising 
evolution channel for CWSs (with $a_\star \gtrsim 0.1$ pc) to attain significant 
eccentricity ($\gtrsim 0.2$) despite their local and resonant-relaxation 
timescale being longer than a few Myr \citep{yu2007}.

With a sufficiently large population of high-eccentricity stars being excited, 
the stage is set for the next phase, dynamical relaxation among the stars 
(\S\ref{sec:dynrelax} and \S\ref{sec:synergy}).  Through resonant relaxation, 
these stars undergo substantial nonlinear reduction of their peri-centric 
distances in combination with a further increase of their eccentricities 
and orbital inclinations, building up the spheroidal S-star cluster.

Thus, as shown in the bottom panels of Figure \ref{fig:a_inc_t}, after $\sim 6-9$ Myr evolution, the eccentricities of most stars formed in the inner region (with $a_0 \sim 0.01$ pc) attains a significant fraction of unity as their periastron distance reduces.  
In this and subsequent figures, all inclinations ($i_{\star}$, $i_{\rm IMC}$) are measured relative to the initial fundamental plane (the initial total angular momentum planes) for the IMC, stars, and their natal gaseous disk.  Although this so-called ``invariant'' plane (total initial-AM plane) evolves with the gas depletion, its initial orientation provides a well-defined set of reference coordinates.

Time-lapse snapshots of the simulated stars' eccentricity $e_\star$ and peri-center 
distance $r_p=(1-e_\star) a_\star$ are shown in Figure \ref{fig:e_peri_a0_t}.
It compares the emergent S-star cluster's eccentricity distribution in the simulation with the observed distribution \citep{burkert2024}.  

Recent observations \citep{burkert2024} of S-stars show a non-uniform eccentricity-pericenter distance distribution ($e_\star-r_p$) with a zone of avoidance. The simulated {\it Fiducial} model reproduces the large spread of the S-stars' $a_\star, e_\star, i_\star$, consistent with the observed pericenter versus eccentricity distributions of the S-stars, including the observed zone of avoidance in the $e_\star-r_{p}$ distribution. At an age of $9$ Myrs the match is excellent.

In the more quantitative Figures \ref{fig:e_peri_a0_t9} and \ref{fig:e_cul_t}, we further compare the $e_\star-r_p$ and cumulative eccentricity distributions of the {\it Fiducial} model with the observed data for the S-stars \citep{burkert2024}, taking into account the detection probability (grey lines in Figure \ref{fig:e_peri_a0_t9}). 
It shows that at an age of $9$ Myrs (comparable to the stellar estimated age and its upper limits), the simulation leads to a dense cluster that matches very well with the observed large spread of the S-stars' kinematic distributions \citep{burkert2024}.

%An agreement is reached on a time scale ($\sim 6-9$ Myr) 
% 
The absence of S-stars with both small perigee and small $e_\star$ 
in the zone of avoidance is a direct consequence of the 
young stars in the Galactic center are formed or captured in one common gaseous disk with marginal gravitational stability from $\sim 0.01$ pc to $\sim 0.2$ pc, and possibly beyond.  
Note that the simulated stars with $r_p < 0.04$ pc (in the S-stars domain) have a wide 
range of initial semi-major axes ($a_0$, color bar in Figure \ref{fig:e_peri_a0_t9}), including some from the outer regions of the disk.  
%SSR and vZLK's influence is accentuated by the wide $a_0$ distribution among stars in the S-star 
%domain (with $r_{\rm p}< 0.04$ pc, Figure \ref{fig:e_peri_a0_t9}).

Even though the {\it Fiducial} model can reproduce the observed data, with regard to the course of future evolution, we anticipate the effect of resonant relaxation will expand into the CWSs and ODSs domains along with stellar evolution in due course.  As a proof of concept, we extended the fiducial model to 20 Myr, by neglecting any changes in $N_\star$ or $M_\star$.
Such a long timescale is needed for any significant change to take place via a
random-walk relaxation process. 

In Figures \ref{fig:e_cul_t} and \ref{fig:i_cul_t_a}, we discuss the cumulative distributions of stellar eccentricity and orbital inclination, respectively, evaluated at evolutionary times of 9 Myr and 20 Myr. 
The normalized cumulative distribution of measurable $e_\star$'s \citep{burkert2024}  
matches observations after $t \sim 6-9$ Myr (Figure \ref{fig:e_cul_t}). 
At 20 Myr (beyond the main-sequence life expectancy 
for the O/WR type S-stars, CWSs, and ODSs), the orbital eccentricity and inclination of S-stars increase further (Figures \ref{fig:e_cul_t} and \ref{fig:i_cul_t_a}) and the disk signature of CWSs evolves towards a uniform inclination ($i_\star$) distribution (Figure \ref{fig:i_cul_t_a}). Eventually, at $\sim$ 20 Myr $\geq \tau_{\rm RR, \varpi}$, dynamical relaxation would erase CWSs' disk kinematic signature. 
The results of these idealized simulations verify 
continued growth of orbital eccentricity and inclination in the CWSs and ODSs, albeit the highly relaxed kinematics of the S-stars is preserved.

In order to compare the {\it Fiducial} model with additional observed data,
we follow the approach of \cite{vonfellenberg2022}, with plotting the distribution of stars' orbital angular momentum vector relative to the invariant plane ($i_\star$, $\Omega_\star - \Omega_{\rm IMC}$), viewed in some arbitrary Sgr A$^\ast$-centric direction in Figure~\ref{fig:cws_ods}. 
In this plotting, four radial regions are considered, including the S-star-dominated region ($a_\star \leq 0.04$ pc), the CWSs' inner region ($0.04~{\rm pc}\leq a_\star \leq 0.08$ pc), the CWSs' outer region ($0.08~{\rm pc}\leq a_\star \leq 0.16$ pc), and the ODSs-dominated region ($0.16~{\rm pc}\leq a_\star \leq 0.5$ pc). Detailed discussion is shown in \S\ref{sec:orientation}.

%It seems that CWSs are surrounded by more ODSs with similar semi-major axis, larger eccentricity
%($e_\star \geq 0.5$), and diffuse inclination.  Sporadic 
%local vZLK effect or SSR mechanism maxima and relaxation-induced $\varpi_\star - \Omega_\star$ 
%mixing lead to the CWS-ODS dichotomy.

In Figure \ref{fig:pomegaomega0}, we also plot stars' relative longitude of periapse ($\varpi_\star - \varpi_{\rm IMC}$) versus relative longitude of ascending node
($\Omega_\star - \Omega_{\rm IMC}$) distribution.  
Resonant relaxation is more effective for S-stars with small semi-major axes ($a_\star \leq 
0.03$ pc). At $t = 9$ Myr, diffuse patches marginally remain in the $\varpi_\star-\varpi_{\rm IMC}$ and $\Omega_\star-\Omega_{\rm IMC}$ distributions, albeit their $e_\star$-$i_\star$ distribution are nearly uniform (top panels in Figure \ref{fig:pomegaomega0}), consistent with the S-star data (marked by grey stars).
For these stars at  $0.03$ pc $\leq a_\star \leq 0.16$ pc, a region dominated by the CWSs, their orbital excitation via resonant relaxation only marginally restrains the vZLK effect and SSR perturbations from the IMC. Consequently, these stars retain their initial kinematic properties and remain concentrated, with similar $\varpi_\star-\Omega_\star$ (lower-right, Figure \ref{fig:pomegaomega0}). 
However, in the IMC's proximity ($a_\star \geq$ 0.16 pc), the vZLK resonance greatly excites stellar inclination with $\varpi_\star - \varpi_{\rm IMC}$ loosely clustered around $\sim 0-\pi/4$ (\S\ref{sec:kozai}. Moreover, their relative longitude of the ascending node $\Omega_\star-\Omega_{\rm IMC}$ spreads out with a nearly uniform distribution between $-\pi$ and $\pi$, which implies that most stars form a torus of ODSs rather than an inclined disk. 

In the top panels of Figure \ref{fig:rp_timescale}, we also plotted the analytic approximation of 
some quantities, including the angular momentum changing timescale and a precession timescale via resonant relaxation ($\tau_{\rm RR, e}$, $\tau_{\rm RR, \varpi}$), the secular interaction timescale ($\tau_{\rm SI}$),  the characteristic vZLK resonance timescale ($\tau_{\rm vZLK}$), the post-Newtonian precession timescale ($\tau_{\rm pN}$), and orbital periods ($\tau_{rm P}$). This plot quantitatively supports the IMC scenario.

In conclusion, although the IMC's direct influence on the S-star domain is relatively weak, their orbits can be excited through their resonant relaxation 
with their surrounding highly eccentric and inclined stars that were scattered by 
the IMC from its neighborhood to their proximity (Figure \ref{fig:scenario}). The 
{\it Fiducial} model is optimized to simulate the S-stars' observed large eccentricity 
and isotropic inclination 
%$i_\star^\prime$ 
within 9 Myr ($\leq \tau_\star$).  This 
rapid excitation from their circular orbits quenches their accretion from 
their natal disk and limits their asymptotic growth to B-type stars. 
The S-stars' zone of avoidance is attainable, provided they were formed 
at $\geq 0.01$ pc from Sgr A$^\star$.
The timescale for resonant relaxation increases with the stars' semi-major axes.
In the region outside the S-star domain, a fraction of the stars retain their innate 
disk signatures with limited excitation of their eccentricities as observed among CWSs. 
Their prolonged retention in their natal disk also enables them to accrete more mass 
and evolving into the observed O/WR stars. In the outskirt of 
the CWSs' domain ($\geq 0.1$ pc), IMC induces a warp with its elevated perturbation 
on the stars relatively close to its inclined orbit 
\citep{lockmann2009, bartko2009, bartko2010, yelda2014}. 
During the gas-disk depletion, the IMC's vZLK effect persists and its secular resonance 
sweeps inward. Along the way, some stars rapidly attain large $e_\star$ and high 
inclination 
%$i_\star^\prime$ 
to form ODSs. Similar to S-stars, the ODSs' asymptotic growth 
is limited to B stars.  Moreover, the IMC's secular transfer of angular momentum with 
the ODSs causes them to episodically cluster around a disk which is nearly orthogonal to 
the CWSs orbital plane \citep{kocsis2011}.

%%%%%%%%%%%%%%%%%%%%%%%%%%%%%%%%%%%%%%%%%%%%%%%%%%%%%%%%%%%%%%%%%%%%%%
%\subsection{Comparison of the {\it Fiducial} model with observations}
\subsection{Eccentricity-excitation mechanisms}
\label{sec:mechanisms}

%Based on our previous extensive model-parameter studies \citep{zheng+2020, zheng+2021}, we 
%present the {\it Fiducial} model and directly compare the simulated results with recent 
%observational data \citep{burkert2024}.

As we qualitatively discussed in \S\ref{sec:dominant}, the young stars' eccentricities
($e_\star$) at the Galactic center are possibly excited by 1) IMC's vZLK resonance
(\S\ref{sec:kozai}), 2) IMC's SSR （\S\ref{sec:kozai}), and 3) gravitational 
interaction between the stars (\S\ref{sec:dynrelax}), depending on their initial 
locations (\S\ref{sec:synergy}). We distinguish three effects based on 
our simulation results.

With the fiducial model (\S\ref{sec:fiducial}), we showed that the observed
eccentricity ($e_{\star}$) excitation in young stars at the Galactic Center can be attributed to three distinct processes, contingent upon their initial locations: i) vZLK resonance driven by the IMC, ii) SSR perturbation from the IMC, and iii) direct gravitational scattering between stars. We use our simulation results to differentiate the regimes where each process predominates.

i) {\it vZLK effect.}
With an IMC orbital inclination of $i_{\rm IMC}^{\prime} = i_{\rm IMC} - i_{\star} = 120^{\circ}$
relative to the initial stellar disk, its secular perturbation via the vZLK resonance is most effective for stars formed near its orbital semi-major axis ($a_{\rm IMC} = 0.35$ pc). For stars with initial semi-major axes $a_0 \sim 0.2$ pc, the vZLK timescale satisfies $\tau_{\rm vZLK} \leq \tau_{\star}$ (Equation \ref{eq:taukozai}, §\ref{sec:kozai}), enabling efficient orbital excitation.
By $t \simeq 1-3$ Myr, a population of stars with $a_0 \sim 0.1-0.2$ pc, is excited to high eccentricities ($e_\star \geq 0.8$), small periastron distances ($r_{\rm p} \ll a_0$, Figure~\ref{fig:e_peri_a0_t}), and inclinations approaching $i_\star \sim \pi/2$ (see the growth in 
$i_\star$ highlighted in the grey shaded region of Figure~\ref{fig:a_inc_t}).
After $3-5$ Myr of evolution, the characteristics of vZLK resonance are evident in both large eccentricity and large inclination excitation for a group of stars (encompassed in the grey shade in the right panel of Figure \ref{fig:a_inc_t} at 5 Myr). 
The semi-major axis of these stars, $a_\star$, changes very little from their preferred original values 
$a_0 \sim 0.2$ pc (see left panel of Figure \ref{fig:a_inc_t}). Consequently, the periastron of these 
excited stars is reduced well inside their initial range (see also in Figure \ref{fig:e_peri_a0_t}).

ii) {\it Sweeping secular resonance (SSR).}
%Since the IMC's orbit is inclined to the gaseous disk by $120^{\circ}$ in the {\it Fiducial} model, it is not subjected to the gas disk potential (\S\ref{sec:gasdisk}). Nevertheless, it oscillates about an ``invariant'' plane due to its gravitational interaction with the disk stars, as shown by the changes in their relative inclination (Figure \ref{fig:a_inc_t}).
%This interaction also induces the IMC to precess. Moreover, the stars precess due to both IMC's secular perturbation, and the gaseous and stellar disk potential. 
Secular resonance occurs when the IMC's and stars' longitude of periapses $\varpi_\star-\varpi_{\rm IMC}$ is maintained at a fixed shift \citep{zheng+2020} when their precession rates match (\S\ref{sec:sweeping}). 
In $3-6$ Myr, the IMC's SSR sweeps (shade in Figure \ref{fig:a_inc_t})
from $0.2 \rightarrow 0.1$ pc (red-to-orange dots, Figure \ref{fig:e_peri_a0_t}) and 
differentially excites $e_\star$.  Through their mutual 
interaction, the IMC and some stars precess with a persistent, 
finite, clumpy $\varpi_\star-\varpi_{\rm IMC}$ distribution
(bottom-right Figure \ref{fig:pomegaomega0}) within the SSR-effective zone 
($a_\star=0.04-0.16$ pc). 
%In the {\it Fiducial} model, IMC's secular resonance also leads to the 
%excitation of some disk stars' (with 0.04 pc $\leq a_\star \leq 0.16$ ~pc) eccentricity towards unity. 
%But the alignment of the longitudes of periapses $\varpi_\star - \varpi_{\rm IMC}$ (bottom right panel %of Figure \ref{fig:pomegaomega0})
%does not significantly change the stars' semi-major axis or inclination.  At 5 Myr, there is a 
%population of stars with high $e_\star$, relatively small inclination (encompassed in the light green 
%shade in the right panel of Figure \ref{fig:a_inc_t}), and limited changes in their $a_\star$ (left panel of Figure \ref{fig:a_inc_t}).
As the gas disk is severely depleted, the inward sweep of IMC's secular resonance is halted by the general relativity (GR) precession and mutual gravitational perturbations between the stars.  
The GR contribution decreases steeply with $a_\star$, whereas the eccentricity excitation timescale due to dynamical relaxation ($\tau_{\rm e}$) increases with $a_\star$ (Equation \ref{eq:taue}).  
In our previous simulations, the GR effect has been taken into account.  But the combined contribution 
of IMC's secular perturbation and the stellar relaxation has not hitherto been simulated and analyzed simultaneously.
After $3-5$ Myr ($1-2~\tau_{\rm dep}$), the relocation of IMC's secular resonance is stalled at $\sim$ $0.1$ pc
where secular perturbation from the IMC is too weak to excite the young stars' orbital eccentricity substantially as the $e_\star$ excitation time becomes too long 
\citep{zheng+2020, zheng+2021}.

iii) {\it Dynamical relaxation.} 
Equation (\ref{eq:taue}) indicates that stellar-disk relaxation is more effective for S-stars with relatively small $a_\star (\leq 0.04$ pc), even though the amplitude of $e_\star$ excitation within 3 Myr is limited (right panel in Figure \ref{fig:a_inc_t}).
Within $t \leq 1$ Myr, it rapidly excites
$e_\star \leq 0.3$ with a modest fractional semi-major axis 
change (Figure 
\ref{fig:e_peri_a0_t}, \S\ref{sec:dynrelax}). 
At $6-9$ Myr, the intruding nearly parabolic stars formed in the outer regions of the disk undergo
resonant relaxation with moderately eccentric stars formed in the inner region (encompassed in the 
patch with light blue shade in the fourth right panel of Figure \ref{fig:a_inc_t}). This interaction 
accelerates $e_\star$, $i_\star$ excitation and leads to phase mixing, randomization of $\varpi_\star$ 
and $\Omega_\star$.  
After $\sim 6$ Myr, it increases
the average eccentricity $\sqrt {\langle e^2 _\star \rangle} \sim 0.5$ with
a comparable semi-major axes altering.  At the same time, 
resonant relaxation (\S\ref{sec:dynrelax}) between the vZLK-SSR-infused stars (with 
$e_\star \sim 1$) and close-in indigenous stars (with $a_0 \sim a_{\rm in}$) randomize $e_\star$, $i_\star$, $\varpi_\star$, $\Omega_\star$, and reduce some S-stars' semi-major axes to $\leq a_{\rm in}$ (Figure \ref{fig:a_inc_t}).
Moreover, although we have neglected the GR correction in this investigation, the stellar relaxation processes are likely to be the dominant effect that suppresses the propagation of IMC's secular resonances. 
Stellar relaxation modifies the stellar semi-major axes $a_\star$ 
from $a_0$ (Figure \ref{fig:a_inc_t}), perturbs $\varpi_\star - \varpi_{\rm IMC}$, 
interrupts the SSR process, and stifles vZLK effect (\S\ref{sec:kozai}).

%At $t= 9$ Myr, diffuse patches marginally remain in the $\varpi_\star-\varpi_{\rm IMC}$ 
%and $\Omega_\star-\Omega_{\rm IMC}$ distributions for the close-in S-stars (the top panel of 
%Figure \ref{fig:pomegaomega0}), albeit their $e_\star$ and $i_\star$ distribution are nearly uniform.
%These results reproduce the limited observational data on the S-stars' kinematic distribution,
%%This expectation reproduces the observed $\varpi_\star-\Omega\star$ distribution among the S-stars.  
%consistent with the S-star data (marked by grey stars). 

%%%%%%%%%%%%%%%%%%%%%%%%%%%%%%%%%%%%%%%%%%%%%%%%%%%%%%%%%%%%%
\subsection{Orientation of stellar orbital planes}
\label{sec:orientation}

Since IMC has a large initial inclination relative to the plane of the natal-disk, these interactions, including the vZKL effect, the SSR perturbation from the IMC, and resonant relaxation among stars, change the orientation of the stellar orbits by varying amounts in various regions. 
 
%We analyze the {\it Fiducial} model, with which we reproduce the observed orbital 
%angular momentum vector distribution for four radial regions following
%the observational approach \citep{vonfellenberg2022}. 
%In Figure \ref{fig:cws_ods}, we characterize, for four radial regions, the 
%distribution of stars' orbital angular momentum vector relative to the invariant plane (total initial-AM plane) and viewed from the SMBH in some arbitrary direction.

%In the region between $\sim 0.04-0.08$~pc, IMC's vZLK and secular-resonance perturbations 
%are relatively weak, and the star's resonant relaxation time exceeds their age.  The $i_\star$ and $e_\star$ distributions clearly show a remnant population that retained 
%their initial kinematic properties. 
%Stars with $a_\star \sim 0.08-0.16$ pc generally avoid close
%ncounters with the IMC, though its cumulative secular perturbation is intense.  Their $i_\star$ distribution also appears to be patchy.  
%Star with $a_\star \sim 0.16-0.5$ pc are scattered by the IMC. 

i) {\it S-stars} with $a_\star \leq 0.04$ pc have a relatively diffuse $i_\star-\omega_\star$ distribution with some notable patches. Under the influence of IMC's vZLK and secular resonant perturbation, they originated from a wide range of initial $a_0$ (Figure \ref{fig:e_peri_a0_t9}). 
Resonant relaxation is more effective for S-stars with small semi-major axes. At their current location, IMC's vZLK and secular-resonance perturbations are relatively weak, S-stars have the shortest dynamical and interaction time scales. With the shortest $\tau_{\rm RR}$, their orbital inclination and orientation have been mostly randomized by resonant relaxation. 
%%% added by zxc
However, a fully isotropic distribution is not achieved in our simulations, which may be due to incomplete relaxation.
%%%

ii) {\it Clockwise disk stars} have been observed with modest $a_\star$, $e_\star$, and $i_\star$.  
We select, from the simulation, populations of inner (0.04 pc $\leq a_\star \leq 0.08$ pc) and 
outer (0.08 pc $\leq a_\star \leq 0.16$ pc) CWSs. At $0.04$ pc $\leq a_\star \leq 0.08$ pc, since the vZLK effect and the SSR perturbation of IMC in this region are modest, and the star's resonant relaxation time exceeds their age ($\tau_{\rm RR} \geq \tau_\star$), many stars retain their innate disk-kinematic signatures with modest $e_\star$ and $i_\star$ (Figure \ref{fig:a_inc_t}). Thus, the inner CWSs are densely concentrated, with a similar orbit-normal direction (upper-right panel of Figure \ref{fig:cws_ods}) and $\varpi_\star-\Omega_\star$ (lower-right panel of Figure \ref{fig:pomegaomega0}). Their prolonged retention in their natal disk may also have enabled CWSs to accrete additional gas and evolve into O/WR stars, more massive than the S-stars and ODSs.

%These stars retain their initial kinematic properties (right, 
%Figure \ref{fig:a_inc_t}) with modest $e_\star$ and $i_\star$. 

%It shows that the inner CWSs' inclination ($i_\star$) and longitude of the ascending node ($\Omega_\star$) are densely concentrated  
%in the upper right panel of Figure \ref{fig:cws_ods}. 
%The resonant relaxation's eccentricity-excitation marginally restrains IMC's vZLK and SSR on some inner CWSs.
%Since the vZLK effect and the SSR perturbation of IMC in this region are modest ($\tau_{\rm RR} \geq \tau_\star$) in this region, many stars retain their innate disk-kinematic signatures.  

iii) A {\it CWSs' warp} have been suggested in the {\it Fiducial} model in the outer region at $0.08$ pc $\leq a_\star \leq 0.16$ pc. 
Within this region, the IMC's vZLK perturbation and SSR are more intense than in regions closer to the SMBH, and generally avoid close encounters with the IMC, these cumulative secular perturbations lead to a patchy $i_\star-(\Omega_\star-\Omega_{\rm IMC})$) distribution (dotted circle in the lower-left of Figure~\ref{fig:cws_ods}), similar to the observed warp structure \citep{lockmann2009, bartko2009, 
bartko2010, yelda2014}. They enhance the growth 
of the warping mode from that resulting from resonant relaxation \citep{kocsis2011}.

%A warped structure similar to that observed \citep{lockmann2009, bartko2009, 
%bartko2010, yelda2014} is indicated by the extended $i_\star-\Omega_\star$
%distribution (dotted circle in the lower-left Figure \ref{fig:cws_ods}) 
%among the outer ($0.08$ pc $\leq a_\star \leq 0.16$ pc ) CWSs.
%Within this region, the orbital eccentricities and inclinations of the CWSs are excited by strong perturbations from the IMC, including vZLK and SSR processes. This excitation, in turn, amplifies the growth of the warping mode driven by resonant relaxation.

%For stars with $a_\star \geq$ 0.16 pc, this results in a loose clustering of $\varpi_\star - \varpi_{\rm IMC}$ around $\sim 0-\pi/4$ whereas the distribution of $\Omega_\star-\Omega_{\rm IMC}$ 
%is less clumpy (bottom-left panel, Figure \ref{fig:pomegaomega0}). This feature
%corresponds to a torus with a high inclination pseudo-disk (Figure \ref{fig:cws_ods}). 

iv) A {\it pseudo counter-clockwise stars disk} \citep{paumard2006, lu2009, yelda2014, ali2020, jia2023} can interpret the observed patchy $i_\star-(\Omega_\star-\Omega_{\rm IMC})$ distribution at $a_\star \geq 0.16$ pc (lower right, Figure \ref{fig:cws_ods}), and a loose clustering of $\varpi_\star - \varpi_{\rm IMC}$ around $\sim 0-\pi/4$ whereas the distribution of $\Omega_\star-\Omega_{\rm IMC}$ 
is less clumpy (bottom-left panel, Figure \ref{fig:pomegaomega0}), which are reproduced by the {\it Fiducial} model. This feature corresponds to a torus with a high inclination pseudo-disk, we refer to this feature as a {\it pseudo disk}, in contrast to the CWSs, which bear some birthmarks of their natal disk. It remains elusive and controversial. They may also be due to the incompleteness of the kinematic data, especially for stars with $a_\star \gtrsim 0.16$ pc.  
Although all stars start on a unique disk plane in our simulation, the intense vZLK effect from the IMC in this region modulates the inclination of ODSs with a comparable maximum amplitude
($\sim i_{\rm IMC}$ in Figure \ref{fig:a_inc_t}) and a dispersed $i_\star-\Omega_\star$ distribution induced by the own resonant relaxation among the ODSs population and the IMC's SSR effect.
Some ODSs' close IMC encounters also introduce large dispersion in their $i_\star-\Omega_\star$ 
distribution.
According to the open circle in Figure \ref{fig:a_inc_t} and the red cross in Figure \ref{fig:cws_ods}, we can find that the orbital angular momentum vector of the IMC ($i_{\rm IMC}-\Omega_{\rm IMC}$) lies in the proximity of the pseudo disk domain, in agreement with the observed kinematic similarity between pseudo counter-clockwise stars and the IRS13E group \citep{jia2023}.  
\subsection{Complementarity of composite effects}
\label{sec:complement}

\begin{figure*}[ht!]
%\centering
\includegraphics[width=0.9\columnwidth]{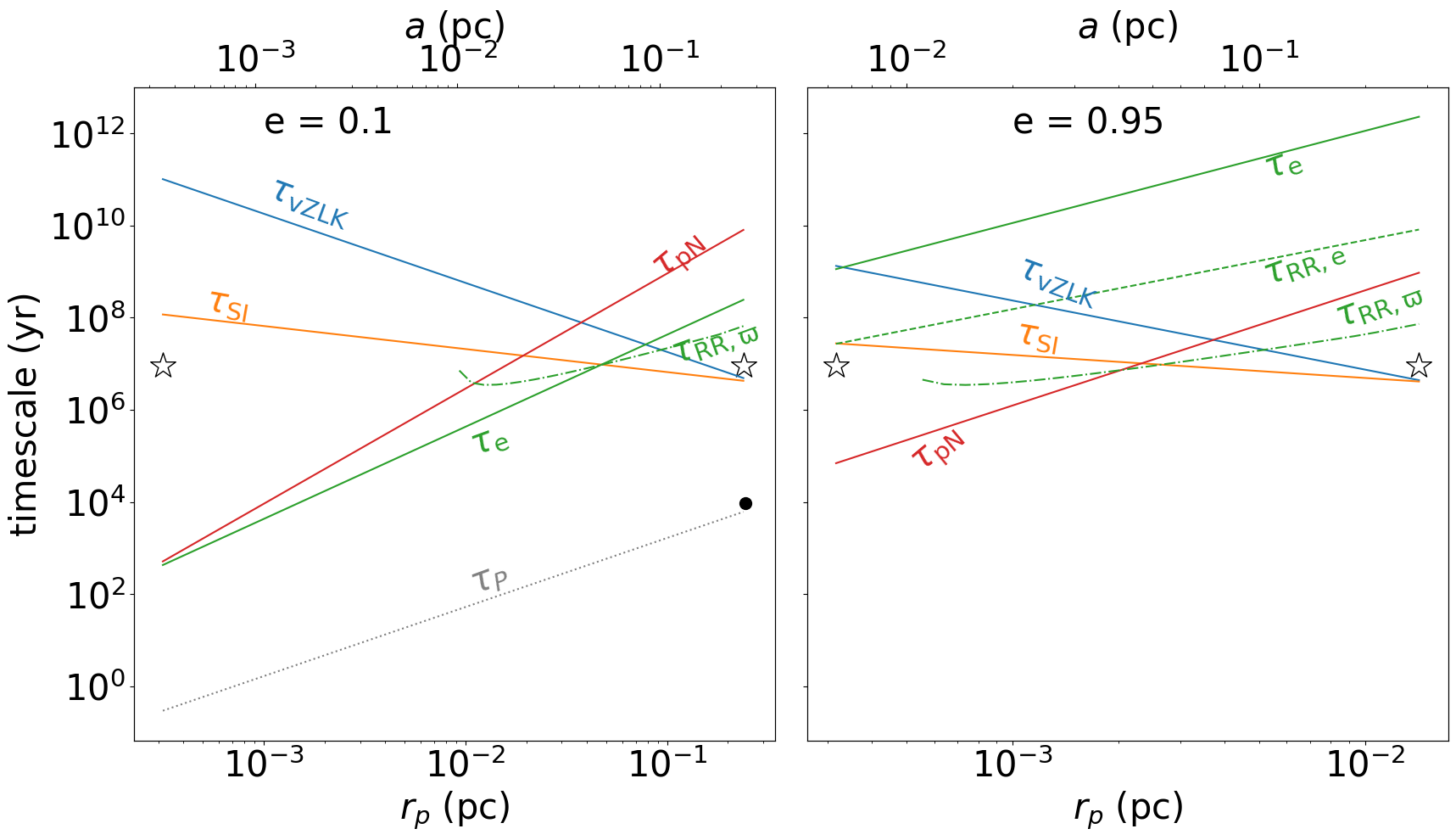}
\includegraphics[width=1.0\columnwidth]{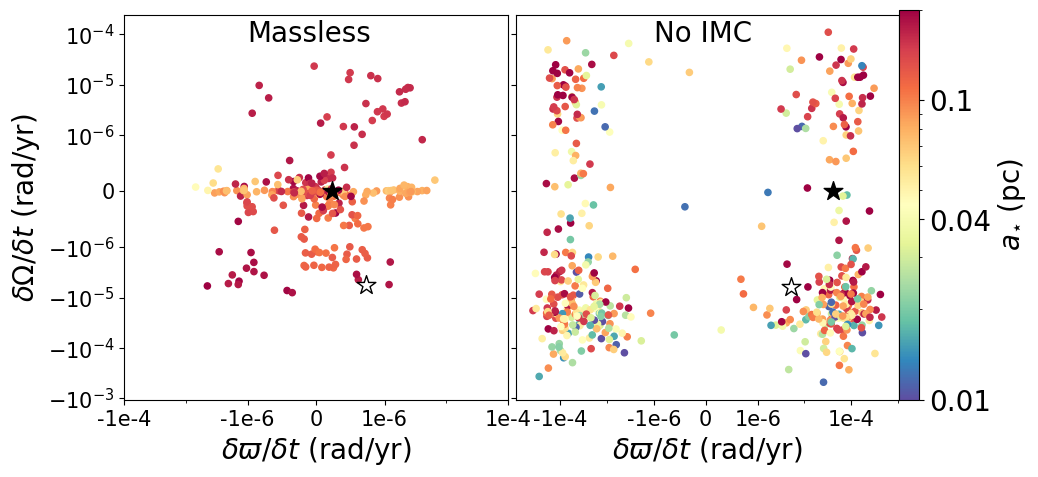}
\caption{ Analytic approximation of
$\tau_{\rm vZLK}$, $\tau_{\rm SI}$, $\tau_{\rm pN}$, 
$\tau_{\rm P}$, $\tau_{\rm e}$, $\tau_{\rm RR, e}$, 
and $\tau_{\rm RR, \varpi}$ as functions of $r_{\rm p}$
for the {\it Fiducial} model with $e_\star= 0.1$ (upper left) and $0.95$ 
(upper right panel). Regions with $a_\star =r_{\rm p}/(1-e_\star) > a_{\rm out}$ 
or $a_\star < a_{\rm in}$ are excluded. Solid dot and open stars denote
IMC's orbital period and stars' age.  With $\tau_{\rm SI} 
\leq \tau_{\rm vZLK} \leq \tau_\star$, 
both vZLK and SSR can excite $e_\star$ of stars with $a_\star \sim a_{\rm IMC}$.
In the S-star domain, the disk stars relaxation can excite $ 0.1 < {\sqrt 
{\langle e_\star ^2\rangle}} = e_\star (\sim 0.5) < 0.95$  
with a comparable fractional change in $a_\star/a_0$ from unity.
Moreover, S-stars' dispersion becomes isotropic 
with $\tau_{\rm RR, \varpi} < \tau_\star$.
Except for a few stars with $r_p \leq 2 \times 10^{-3}$pc, 
$\tau_{\rm SI} \leq \tau_{\rm vZLK} \leq \tau_{\rm pN}$
i.e. precession due to IMC's secular perturbation 
is comparable to or larger than the Schwarzschild precession. 
Over multiple periods, the vZLK effects leads 
to changes in $i_\star$, i.e. precessions in directions generally off 
the orbital plane.  These approximations
reproduce the longitudinal ($\delta \varpi / \delta t$) and nodal 
($\delta \Omega / \delta t$) precession rates  
of stars with $e_{\star} > 0.05$ after 9 Myr evolution
for the {\it Massless} (lower left) and the {\it No IMC} 
(lower right) models.
The maximum post-processed GR precession rates (solid stars) 
is much smaller than their values (hollow stars) solely due 
to IMC's perturbation in the {\it Massless} model where 
large $e_\star$'s are only excited among stars in the IMC's 
neighborhood. The comparable magnitudes between the hollow and solid stars for the {\it No IMC} model suggest the GR precession does not suppress most stars' relaxation.
}
\label{fig:rp_timescale}
\end{figure*}

\begin{figure*}
\centering
 \includegraphics[width=0.75\columnwidth]{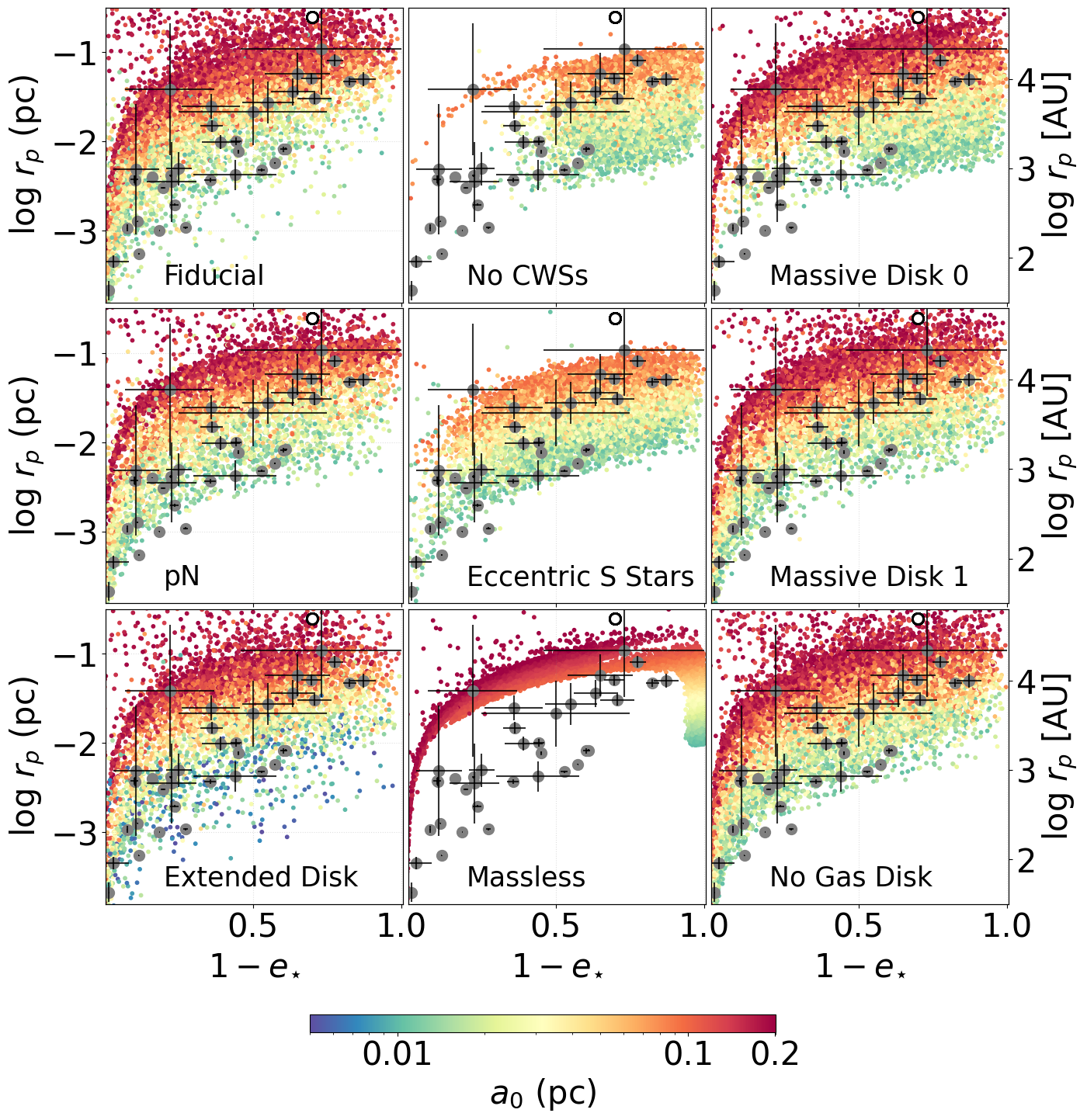}
 \caption{The perigee versus $1-e_\star$ distribution 
of observed S-stars (black with error-bars) and simulated stars 
as a function of their initial location 
(color) for various idealized models at $t=9$ Myr.
They highlight contributions from IMC's vZLK and 
SSI effects as well stellar relaxation.}
 \label{fig:e_peri_a0_t9_mc}
\end{figure*}

\begin{figure*}
\centering
 \includegraphics[width=0.75\columnwidth]{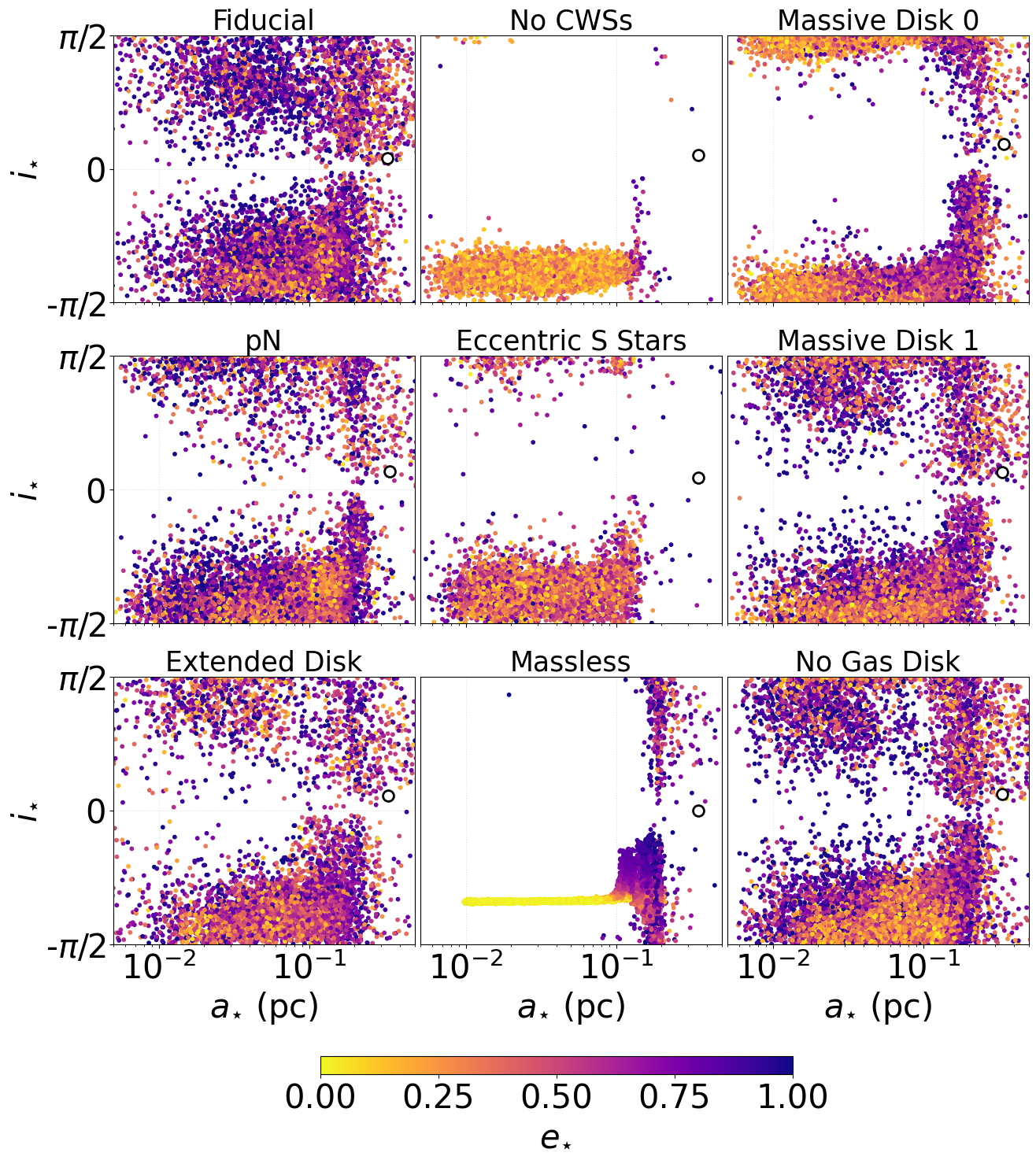}
 \caption{Stars' and IMC's (open circle) $a_\star-i_\star$ (relative to the I-plane) distribution and 
 $e_\star$ (color) at $t=9$ Myr for various models. 
 Some stars in the {\it Fiducial} models are scattered, by the IMC, to orbits with $a_\star \geq 
 a_{\rm out}$, including some with $i_\star$ similar to the CWSs.}
 \label{fig:a_inc_e_t9_mc}
\end{figure*}

\begin{figure*}
\includegraphics[width=1\columnwidth]{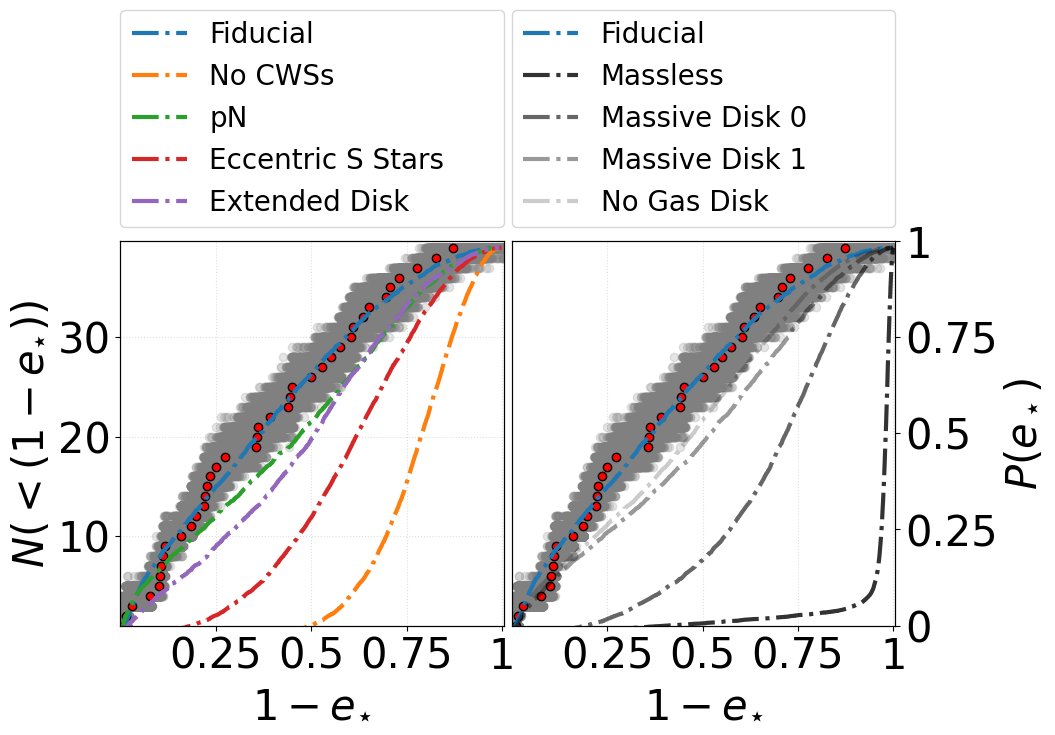}
 \caption{% $P(e_\star)$  
The normalized cumulative $e_\star$-distribution of all the simulated stars, including detection probability, 
at $t=9$ Myr for various idealized models. The red dots and grey shades are defined in Figure~\ref{fig:e_cul_t}. Close match between {\it Fiducial} model with observed data highlights the
combined influence of IMC's vZLK and SSR perturbation and the stars' resonant relaxation.
 }
 \label{fig:e_cul_mc}
\end{figure*}

With additional simulations (variations of the {\it Fiducial} model), 
we identify dominant physical processes, analyze 
the dynamical evolution of various stellar populations, and place constraints on the 
IMC's orbit and stellar properties in the method section. These idealized models show
the necessity of an IMC on a highly inclined orbit with respect to the initial
stellar disk, in reproducing the S-stars' high eccentricity-inclination
%$i_\star ^\prime$ 
today.  
We subsequently verify these model parameters to confirm and highlight the dominant physical effects responsible for the whole evolution.

The {\it Fiducial} model (upper left panel, Figures \ref{fig:e_peri_a0_t9_mc} \& \ref{fig:a_inc_e_t9_mc}) incorporates all three separate eccentricity and inclination ($e_\star-i_\star$) excitation mechanisms. 
All of these effects involve various types of secular interaction which, to first order, do not significantly modify the stars' semi-major axes $a_\star$ from their initial values $a_0$ (see also in the Figure \ref{fig:a_inc_t}). 
We analyze their interdependence by
suppressing various contributions (Figures 
\ref{fig:e_peri_a0_t9_mc}, \ref{fig:a_inc_e_t9_mc}, 
\ref{fig:e_cul_mc}) with additional sets of idealized 
simulations for the variational models. 
The approximate formulae in \S\ref{sec:kozai}
$-$\ref{sec:dynrelax} agree well with those extrapolated from the idealized 
{\it Massless} and {\it No IMC} models (bottom panels Fig. \ref{fig:rp_timescale}).

\noindent
A) The {\it Massless} model (upper-right panel, Figures \ref{fig:e_peri_a0_t9_mc} \& \ref{fig:a_inc_e_t9_mc}) suppresses the effects of disk and resonant relaxation of stars
by setting $M_\star=0.1~M_{\odot}$.
For stars with initial semi-major axes $a_0 > 0.1$ pc, the vZLK effect excites orbits high eccentricity  $e_\star$ and high inclination $i_\ast$, while the SSR process 
induces high eccentricity with low inclination. 
Without resonant relaxation, stars with $a_0 \leq 0.1$ pc can retain their initial eccentricity and inclination, so that $e_\star \sim e_0$ and $i_\star \sim i_0$.
In the upper right panel of Figure \ref{fig:e_peri_a0_t9_mc}, a direct comparison of the $(1-e_\star)-{r_p}$ distribution of this model with that of detectable stars in the {\it Fiducial} model (Figure \ref{fig:e_peri_a0_t9}) highlights the need for stellar relaxation in reproducing the observed distribution for the S-stars (see also in Figure \ref{fig:e_cul_mc}).  
Among the high-$e_\star$ outer stars ($a_0 > 0.1$ pc), the population with large $i_\ast$ is excited by the vZLK resonance, whereas that with low $i_\star$ is excited by the SSR process (upper right panel, Figure \ref{fig:a_inc_e_t9_mc}).
Although high-$e_\star$ outer stars are produced in the {\it Massless} model, their eccentricity distribution does not match the observational data (Figure \ref{fig:e_cul_mc}). 
%The $e_\star$-distribution does not match the observed data
%(Figure \ref{fig:e_cul_mc}).

%This result confirms the suggestion that although disk relaxation \citep{palmer1993, aarseth1993, yelda2014}
%can excite modest orbital eccentricities ($e_\star \sim 0.2$), resonant relaxation alone \citep{hopman2006, kocsis2011} cannot 
%excite 
%

\noindent 
B) The {\it No CWSs} model 
(upper-middle panel, Figures \ref{fig:e_peri_a0_t9_mc} \& \ref{fig:a_inc_e_t9_mc}) reduces the vZLK resonance and delays the SSR process (Figure \ref{fig:rp_timescale}) by confining all stars to form 
with $a_0 = 0.01-0.1$ pc and initial circular orbits ($e_0=0$). This configuration ensures the vZLK timescale exceeds the stellar age ($\tau_{\rm vZLK} > \tau_\star$, Equation \ref{eq:taukozai}).
After the 9 Myr simulation, most stars attain only modest eccentricities ($e_\star \lesssim 0.5$), albeit
below the observed values (see the upper middle panel of Figure \ref{fig:e_peri_a0_t9_mc} and the orange dash-dotted line in Figure \ref{fig:e_cul_mc}).
This result confirms the suggestion that although disk relaxation \citep{palmer1993, aarseth1993, yelda2014}
can excite modest orbital eccentricities ($e_\star \sim 0.2-0.5$) with roughly initial inclinations ($i_\star \sim i_0$), resonant relaxation alone \citep{hopman2006, kocsis2011} cannot effectively excite the large observed $e_\star$ within a few Myr (\S\ref{sec:dynrelax}).
A direct comparison between the lower right panel of Figures \ref{fig:a_inc_t} and the upper middle panel of Figure \ref{fig:a_inc_e_t9_mc} also indicates that on the time scale of 9 Myr, the vZLK 
resonance is more effective in exciting $i_\star$ for the ODSs and warping the disk for
the CWSs than the resonant relaxation process \citep{kocsis2015}.

\noindent
C) The {\it Eccentric S-stars} model (middle panel, Figures \ref{fig:e_peri_a0_t9_mc} \& \ref{fig:a_inc_e_t9_mc}) minimizes the IMC's vZLK resonance and SSR perturbation  
with a population of eccentric stars formed in a well-confined region $a_0 = 0.01-0.1$ pc and imposes a large initial angular momentum deficit (AMD) with $e_0=0.5$ following
the eccentric-stream scenario
\citep{Alexander2007, BonnellRice2008, HobbsNayakshin2009}. 
Similar to the {\it No CWSs} model, the effects of vZLK 
resonance is suppressed by setting a confined initial semi-major axes with $a_0 = 0.01-0.1$ pc.
Although with finite initial AMD, resonant relaxation alone can boost some stars' eccentricity to be comparable to that of the S-stars (middle left panel of Figure \ref{fig:e_peri_a0_t9_mc}). 
Comparison between this model (middle left panel of Figure \ref{fig:a_inc_e_t9_mc}) and the {\it Fiducial} model (Figure \ref{fig:a_inc_t}) indicates in the absence of 
vZLK resonance, the inclination of this population does not increase significantly. 
The S-stars' observed isotropic $i_\star-e_\star$
and cumulative $e_\star$ distributions cannot be reproduced (see also in Fig \ref{fig:e_cul_mc}).
 
\noindent
D) The {\it No Gas Disk} model (lower-right panel, Figures \ref{fig:e_peri_a0_t9_mc} \& \ref{fig:a_inc_e_t9_mc}) 
suppresses the effect of the SSR perturbation  
by neglecting the gas disk's contribution to the 
potential, i.e., by setting $\Phi_{\rm \star, disk} 
= \Phi_{\rm \star, IMC} = 0$ (Equations \ref{eq:f_star} \& 
\ref{eq:f_IMC}).
The middle panel of Figure~\ref{fig:e_peri_a0_t9_mc} and the middle panel of Figure~\ref{fig:a_inc_e_t9_mc} show that the simulated S-stars can attain significant 
orbital excitation both in the eccentricity and inclination within 9 Myrs under IMC's vZLK resonance and stars' resonant relaxation, even without the continuous 
influence of the IMC's SSR perturbation.
However, although the stellar orbits ($e_\star-i_\star$) are excited
under IMC's vZLK resonance and stars' resonant relaxation, compared to the S-stars' observed isotopic 
$i_\star-e_\star$ distribution, some of these simulated S-stars in the {\it No Gas Disk} model concentrate at the same plane with low eccentricity, their magnitudes are insufficient to match the observed
$e_\star$-distribution (Figure \ref{fig:e_cul_mc}).
It indicates the insufficient eccentricity excitation of young 
stars under the combined effect of the vZLK resonance and resonant relaxation. Negligence of
the SSR prolongs the time required for the cumulative $1-e_\star$ distribution to
reach its observed values (Figure \ref{fig:e_cul_mc}).

\noindent
E) The {\it Massive Disk} models 0 (with $\tau_{\rm dep} = 2.5$ Myr, 
top-right panel, Figures \ref{fig:e_peri_a0_t9_mc} \& \ref{fig:a_inc_e_t9_mc}) and 1 (with $\tau_{\rm dep}=1$ Myr, middle-right panel, Figures \ref{fig:e_peri_a0_t9_mc} \& \ref{fig:a_inc_e_t9_mc})
weakens the vZLK resonance and slows down the secular resonance's sweeping propagation.
The total initial mass of the gaseous disk $M_{\rm disk}$ in the {\it Fiducial} model is slightly 
comparable to that of the IMC and is slightly larger than the stars' total mass (\S\ref{sec:gasdisk}).
In the {\it Massive Disk} models 0 and 1, the gaseous disk is an order of magnitude larger $\Sigma_0$ and $M_{\rm disk}$, with $\Sigma_0=8 \times 10^3$ g cm$^{-2}$ and $\tau_{\rm dep} = 2.5$ and $1$ Myr respectively (Equation \ref{eq:sigma}).   
After $9$ Myr, the cumulative $e_\star$-distributions in model 0 
do not converge onto the observed values (right panel of Figure \ref{fig:e_cul_mc}).
Nevertheless, a rapid depletion of the disk gas (with a relatively short $\tau_{\rm dep}$) does suppress 
these stabilizing effects. 
%though rapid disk depletion (model 1) reduces these discrepancies. 
The asymptotic outcome of the star's eccentricity and inclination excitation depends more sensitively on the depletion time scale of the gaseous disk ($\tau_{\rm dep}$) rather than on the radial distribution $\Sigma_{\rm g}$.

\noindent
F) The {\it pN} model (middle-left panel, Figures \ref{fig:e_peri_a0_t9_mc} \& \ref{fig:a_inc_e_t9_mc})
with post-Newtonian corrections does not statistically modify stars' dynamical evolution.
Although the Schwarzschild precession quenches the SSR perturbation of the IMC very close to the SMBH, it is negligible for most S-stars (Figure \ref{fig:rp_timescale}, \ref{fig:a_inc_e_t9_mc}, \& \ref{fig:e_cul_mc}). 
For most S-stars with the perigee distance to the \sgra follows $r_{\rm p} \geq r_\bullet M_\bullet/ 2 
M_\star \sim 0.01$ pc (Figure \ref{fig:e_peri_a0_t9}). 
The Schwarzschild precession (over timescale $\tau_{\rm pN} 
\simeq P_\star (1+e_\star)r_{\rm p}/r_\bullet$) 
\citep{levin2007, lu2007, misner1973, hopman2006} has been directly observed in the orbit of star S2 \citep{gravity2020}. And post-Newtonian corrections weakly affect tidal 
disruption and hyper-velocity ejection of stars 
\citep{iorio2020, peissker2020,fragione2020, levin2003, rantala2024}. 
Stars in our scenario begin with low eccentricities and inclinations  ($e_0 \simeq i_0 ^\prime \simeq 0$),
and semi-major axis $a_{\rm in} < a_0 \leq a_{\rm out}$. For these orbits $a_0 \geq 0.02$ pc, the post-Newtonian corrections is weaker than the secular interaction of the IMC as $\tau_{\rm SI} \leq \tau_{\rm pN}$. For those stars at $a_0 \geq 0.05$ pc, the orbit precession induced by the post-Newtonian effect can be ignored as $\tau_{\rm vZLK} \leq \tau_{\rm pN}$ (top, left Figure \ref{fig:rp_timescale}).
Importantly, for the present-day S-star population, resonant relaxation in the longitude of periastron operates faster than post-Newtonian precession ($\tau_{\rm RR, \varpi} < \tau_{\rm 
pN}$, top right Figure \ref{fig:rp_timescale}).
Consistent with this analysis, our test models confirm that 
Schwarzschild precession neither suppress stellar relaxation (bottom right panel of Figure \ref{fig:rp_timescale}), 
detune the vZLK effect and SSR mechanism of the IMC (bottom left panel of Figure \ref{fig:rp_timescale})
nor statistically alter the eccentricity excitation pathways for most S-stars, CWSs, and ODSs (Figures \ref{fig:a_inc_e_t9_mc} \& \ref{fig:e_cul_mc}).

\noindent
G) The {\it Extended Disk} model (lower-left panel, Figures \ref{fig:e_peri_a0_t9_mc} \& \ref{fig:a_inc_e_t9_mc}) 
with $a_{\rm in}=0.005$ pc, i.e. initial $a_\star$ in the same 
range ($0.005-0.02$ pc) as its present-day values. 
The diffusion of stars into the zone of 
avoidance within $9$ Myr in the {\it Extended Disk} model (lower-left) of Figure
\ref{fig:e_peri_a0_t9_mc} (cf the {\it Fiducial} model) suggest a more distant star forming region
(with $a_\star \geq a_{\rm in}$, \S\ref{sec:diskstars}).

\begin{figure*}
\centering
 \includegraphics[width=1\columnwidth]{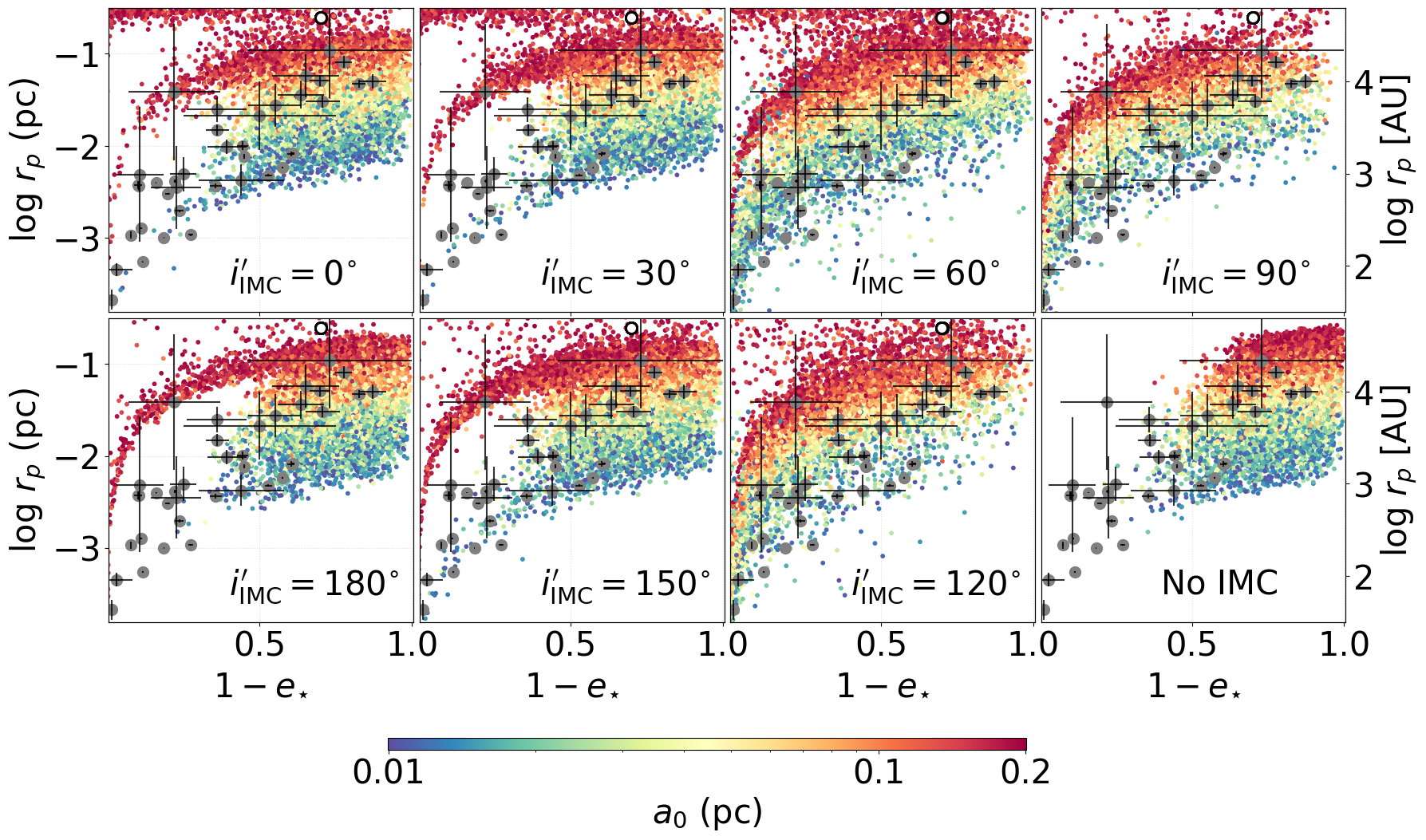}
 \caption{The simulated distribution of various inclination models at $9$ Myr, with the same symbols as Figure~\ref{fig:e_peri_a0_t}. IMC's vZLK effect leads to the rapid $e_\star-i_\star^\prime$
 excitation for $i_{\rm IMC}=60^{\circ}, 90^{\circ}, 120^{\circ}$ and it is quenched at $i_{\rm IMC}=0^{\circ}, 30^{\circ}, 150^{\circ}, 180^{\circ}$. The {\it No IMC} model indicates that stellar relaxation alone cannot excite $e_\star$ to the observed values within 9 Myr.}  
\label{fig:e_peri_a0_t9_inc}
\end{figure*}

\noindent
H) The {\it No IMC} model eliminates IMC's vZLK resonance and SSR perturbation with assuming a massless IMC with $M_{\rm IMC}=0$. 
Without IMC-induced angular momentum deficit infusion, the combined effects of 
star-disk relaxation (\S\ref{sec:dynrelax}) and subsequent 
resonant relaxation are limited. Consequently, stellar eccentricities near the inner disk $a_{\rm in}$ are curbed to only $e_\star \leq 0.3-0.5$
(bottom-right, Figure \ref{fig:e_peri_a0_t9_inc}), well 
below S-stars' observed values.

\begin{figure*}[ht!]
%\centering
\includegraphics[width=1.0\columnwidth]{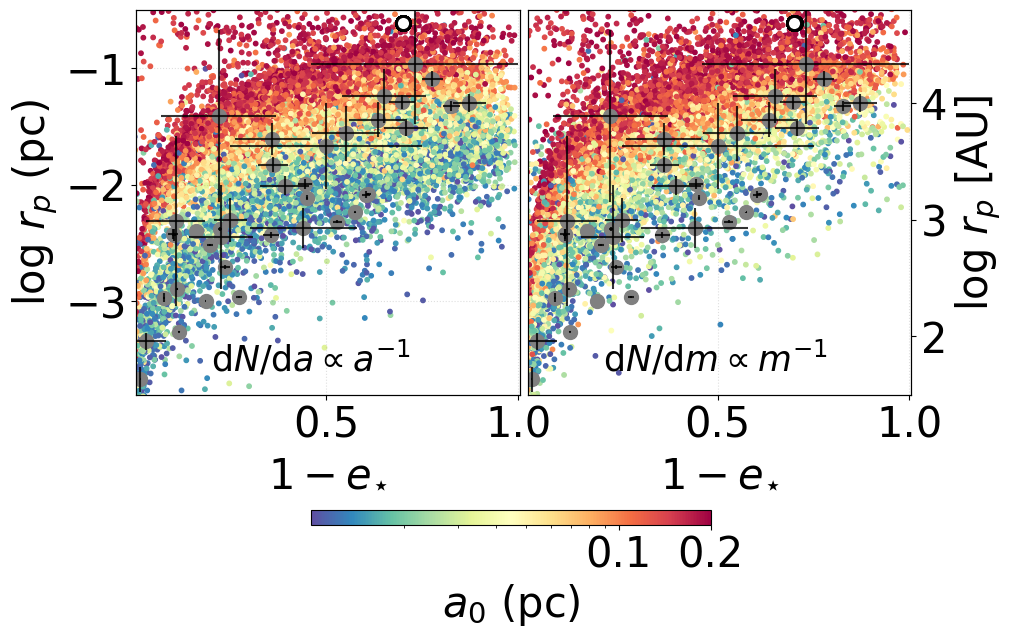}
\caption{The $e_{\star}-r_{\rm p}$ distribution 
at $\tau_\star = 9$ Myr for the {\it Steep} (left) and Kroupa IMF ({\it IMF}, right) model.
There is no significant dependence on the IMF.}
\label{fig:IMF}
\end{figure*}

\noindent 
I) The {\it Steep} and {\it IMF} models adopt a steep stellar surface density profile ($S_\star \propto a_0^{-2}$) and a Kroupa initial mass function (IMF) \citep{marks2012}, respectively. While both models are constructed to have the same total stellar mass and number of stars $N_\star$ as the fiducial model (see Table \ref{tab:models} for parameters), they yield results similar to it. As shown in the left and right panels of Figure \ref{fig:IMF} for the {\it Steep} and {\it IMF} models, respectively, both produce eccentricity distributions comparable to the fiducial model (bottom right panel of Figure \ref{fig:e_peri_a0_t}).

\begin{figure*}
\centering
 \includegraphics[width=0.75\columnwidth]{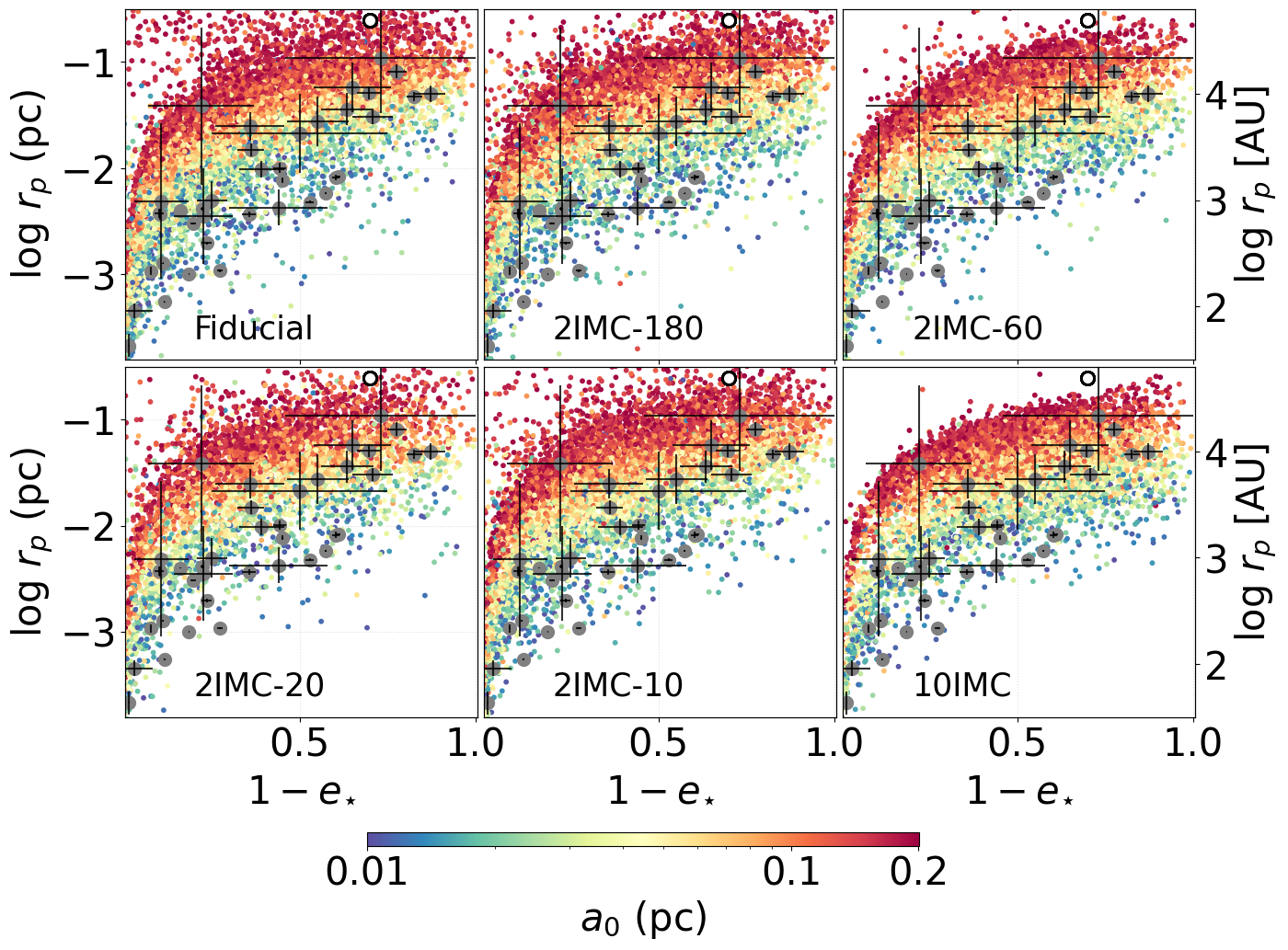}
 \caption{The simulated distribution, at 9 Myr, of models with multiple IMC's on the 
 same orbit with a range of phase separations. Same symbols are used as in Figure~\ref{fig:e_peri_a0_t}. Similarities between these multiple-IMC models 
 characterize the accumulative nature of IMC's secular distant perturbation.
 }  
 \label{fig:e_peri_a0_t9_IMC}
\end{figure*}

\noindent
J) The {\it 2IMC-10}, {\it 2IMC-20}, {\it 2IMC-60}, {\it 2IMC-180} , and {\it 10IMC} models consider the possibility of the intermediate-mass moving groups rather than an IMBH. These models produce very similar $r_{\rm p}-e_{\star}$ distributions for most S-stars and CWSs candidates (Figure \ref{fig:e_peri_a0_t9_IMC}),
validate the orbit-averaging approximation for the vZLK reosnance and the SSR perturbation  \citep{zheng+2021}. 
Nevertheless, the excitation of nearly-escaping orbits (with $e_\star \sim 1$, large $r_{\rm p}$ and $a_\star$) requires some energy changes due to close encounters and therefore it is less effective in Model {\it 10IMC}.

%%%%%%%%%%%%%%%%%%%%%%%%%%%%%%%%%%%%%%%%%%%%%%%%%%%%%
\subsection{Dependence on IMC's inclination}
\label{sec:imcinclination}

The orbital parameters of the IMC in our {\it Fiducial} model are adopted from the 
previous simulations \citep{zheng+2020, zheng+2021}. 
While the onset of vZLK resonance requires an IMC inclination $i_{\rm IMC} \gtrsim 40^{\circ}$ (\S\ref{sec:kozai}), the influence of SSR does not depend on $i_{\rm IMC}$ (\S\ref{sec:sweeping}).
We re-examine the relative contributions of these two effects by conducting a series of models identical in all parameters except $i_{\rm IMC}$.
%the range of IMC's orbital configurations which can reproduce
%the observed kinematic distribution of the stellar populations around Sgr 
%A$^\star$.  
Four of these models {\it INC-0}, {\it INC-30}, {\it INC-150} and {\it INC-180}
with $i_{\rm IMC} ^\prime =0, \pi/6, 5 \pi/6,$ $\pi$ 
do not satisfy the $40^{\circ} \leq i_{\rm IMC}^\prime 
 \leq 140^{\circ}$ vZLK-resonance criterion (\S\ref{sec:kozai}). 
 In these, the SSR process alone excites some stellar eccentricities but fails to reproduce the observed pericenter–eccentricity ($r_p-e_\star$) distribution (Figure \ref{fig:e_peri_a0_t9_inc}). In contrast, the vZLK-effective ({\it INC-60}, {\it INC-90} and {\it INC-120}) models with $i_{\rm IMC} ^\prime =\pi/3, 
\pi/2,$ $2\pi/3$ in Figure \ref{fig:e_peri_a0_t9_inc}, indicate that at 9 Myr:
% But it is expected to dominate three other models in which
%$i_{\rm IMC} ^\prime =\pi/3, \pi/2,$ and $2\pi/3$.
 %While the SSR process
%excites some stars' $e_\star$ (\S\ref{sec:sweeping}), it alone
%does not lead to the observed $r_{\rm p}-e_\star$-distribution
%(Figure \ref{fig:e_peri_a0_t9_inc}).  But, 
%the $r_{\rm p}-e_\star$-distribution of the  
%vZLK-effective models (with $i_{\rm IMC} ^\prime =\pi/3, 
%\pi/2,$ $2\pi/3$) at 9 Myr in Figure \ref{fig:e_peri_a0_t9_inc} indicate that:

\begin{itemize}
    
\item 
Provided that the domain of star formation reaches the proximity ($\sim 0.2$ pc) of the IMC's orbit,  high $e_\star$ can be excited by both vZLK and sweeping secular resonances;

\item
The IMC's SSR alone is less effective in inducing nearly parabolic stars to attain close encounters with Sgr A$^\star$; and

\item 
a sufficiently large population of intruding, nearly parabolic stars is needed to enhance resonant relaxation with stars formed in the inner region of the gas disk.

\end{itemize}

Based on the accumulative $(1-e_\star)$ distribution in the Figure \ref{fig:e_cul_mc}), the IMC in the {\it Fiducial} model appears to provide:

\begin{itemize}
\item 
The most efficient excitation of high $e_\star$ and $i_\star$ (relative to the initial disk) for the ODSs; 

\item
The retention of modest $e_\star$ and $i_\star$ by a fraction of stars 
in the region $0.05$ pc $\leq a_0 \leq$ 0.2 pc, analogous to the CWSs; 

\item
The prolific infusion of nearly parabolic ($e_\star \sim 1$) and high $i_\star$ stars into the S-stars domain;  

\item 
The rapid (within $\leq 9$ Myr) establishment of nearly isotropic velocity dispersion for the S-stars close to the Sgr A$^\star$ (Figures \ref{fig:a_inc_t} and \ref{fig:e_peri_a0_t9_inc}). 

\end{itemize}

%%%%%%%%%%%%%%%%%%%%%%%%%%%%%%%%%%%
%\subsection{Intermediate-mass moving groups}

%With the Lagrange-Laplace orbit-averaging approach \citep{murray2000}, a single IMC's secular perturbation is equivalent to multiple sub-systems with the same total mass and orbital elements 
%($\varpi_{\rm IMC}, \Omega_{\rm IMC}$) at random phases. These sub-systems may represent the tidally-disrupted relics of a parent star cluster. We simulate models {\it 2IMC-180} (\S\ref{sec:complement}G), {\it 2IMC-60}, {\it 2IMC-20}, {\it 2IMC-10} with two IMC with equal mass $0.5\times10^4 M_{\odot}$ and various face angles ($\theta$), $\Delta \theta = 180^{\circ}, 60^{\circ}, 20^{\circ}$ and $10^{\circ}$, respectively. We also consider ten $10^3 M_{\odot}$ IMC with equal initial phase separations on the same orbit in Model {\it 10IMC}. 

%\iffalse
%\noindent
%{\color {blue}
%G) Similar $e_\star-i_\star$ distributions 
%between the {\it Fiducial} and {\it two-IMC} models,
%(lower-left, each with $M_{\rm IMC} =5 \times 10^3 
%M_\odot$ on the same orbits with $180^{\circ}$ or 
%other phase separation), validate the orbit-averaging approximation for vZLK and SSR \citep{zheng+2021}.}
%\fi

%%%%%%%%%%%%%%%%%%%%%%%%%%%%%%%%%%%%%%%%%%%%%%%%%%%%%%%%%%%%%%%
\section{Summary and discussions}
\label{sec:summary}
The Galactic center is an invaluable platform for studying stellar dynamics. 
It contains an SMBH surrounded by a group of B stars within 0.05 pc, commonly referred to as the S-stars, a population of O/WR stars at more considerable distances (out to 0.2 pc), commonly referred to as the clockwise disk stars, CWSs, and another group of ``off-disk'' stars (ODSs) surrounding the CWSs and has a similar luminosity function as the S-stars.
These three populations have similar ages ($6 \pm 2$ Myr and $\lesssim 15$ Myr).  The S-stars have nearly isotropic phase space distribution, including some with $e_\star$ close to unity.  The ``disk'' stars (CWSs) orbit around Sgr A$^\star$~in a relatively thin disk with modest $e_\star (\lesssim 0.4)$ and a warp.  The ODSs are located at $\sim 0.05-0.5$ pc from Sgr A$^\star$ with significantly high $e_\star (\gtrsim 0.4)$, $i_\star$, but non-isotropic distribution, and may or may not have a systematic counter-clockwise rotation around the Sgr A$^\star$  \citep{yelda2014, vonfellenberg2022}.

This discord suggests that although these three populations of young stars may have originated in a common natal disk around Sgr A$^\star$, they may have undergone disparate evolutionary paths, resulting in the distinct present-day orbital distribution between S-stars, CWSs, and ODSs.
The main challenge is attaining such a complex dynamical structure for all these populations of young stars in an {\it unified and self-consistent} scenario.

For computational simplicity, we a) adopted a population of coeval, single stars 
with identical masses and no evolution, b) neglected dynamical relaxation with 
individual low-mass mature, main-sequence stars in the same neighborhood, 
c) only considered the gas contribution to the axisymmetric gravitational potential, and d) neglected relativistic corrections.

%%%%%%%%%%%%%%%%%%%%%%%%%%%%%%%%%%%%%%%%%%%%%
%\subsection{Dominant physical processes}

We assume these stars formed with nearly circular Keplerian orbits in a common natal 
marginally self-gravitating, nearly-axisymmetric, gaseous disk.  Neighboring stars have relatively small
differential azimuthal velocities and long synodic periods.  Through stellar-disk 
relaxation, these stars' eccentricity grows on a time scale $\tau_{\rm e} \propto 
\langle e_\star ^2 \rangle^2$ \citep{palmer1993, aarseth1993} and to the magnitude 
($\lesssim 0.2$) much smaller than those observed values among the S-stars.  

Under the dominant gravitational field of the SMBH, the dynamical evolution of these stars with 
modest eccentricity is different from that in typical stellar clusters.  With small departures from closed
Keplerian orbits, these stars undergo resonant relaxation through mutual secular perturbation 
\citep{rauch1996, kocsis2011, kocsis2015}. Nevertheless, dynamical relaxation among these S-stars, CWSs, 
and ODSs alone are not able to elevate the magnitude of $e_\star$ to be comparable to that observed 
for the S-stars and ODSs (top middle panel of Figure \ref{fig:e_peri_a0_t9_mc}) on a time scale 
comparable to their young ages \citep{hopman2006}.

In our previous investigations, we explored the possible dynamical influence of a potential 
intermediate-mass companion (IMC), e.g., IRS 13E. We conducted extensive surveys to study its 
secular perturbation on the young stars during and after the depletion of their natal disk.  
Our past investigation focused on the evolution of massless particles under the influence 
of IMC's secular resonance as it sweeps inwards from the IMC's proximity 
over an extended region.  These simulations also included the effect of 
vZLK resonance for an IMC with a highly inclined orbit relative to the disk plane.

Based on the previous survey, we chose a {\it Fiducial} model.  As a natural extension of our
previous investigations, we include the dynamic interaction between young stars.
We show that within the estimated age ($6 \pm 2$ Myr and $\lesssim 15$ Myr) of these young stars, 
\begin{itemize}
\item
disk and resonant relaxation among the S-stars can lead to 
the rapid excitation of significant but not nearly parabolic eccentricity
(top middle panel of Figure \ref{fig:e_peri_a0_t9_mc});

\item
in the absence of resonant relaxation, IMC's vZLK resonance and sweeping secular 
resonance alone can induce nearly parabolic eccentricities with a limited inclination 
for most and a large inclination for some ODSs (middle panel of Figure 
\ref{fig:a_inc_e_t9_mc});  and  

\item perturbation by the intruding disk stars 
can promote disk and resonant relaxation, eccentricity excitation, and inclination growth for the S 
stars (bottom right panel of Figure \ref{fig:a_inc_t}). 
Moreover, we show here the complementarity between these effects 
enhances the rapid dynamical relaxation of the young stars in the proximity of Sgr A$^\star$ (Figure \ref{fig:e_cul_mc}).

\end{itemize}

%\end{item}

%\subsection{Luminosity function and kinematics}
%\label{sec:formation}

In an attempt to reproduce their most recently observed $e_\star$ and $i_\star$ distributions 
as well as luminosity function, 
we examine the relative impacts of IMC's vZLK and secular resonances, as well as the 
intrinsic dynamical interaction between these young stars.
Based on the computational results and the analysis of the {\it Fiducial} model 
(\S\ref{sec:results}, Figures \ref{fig:a_inc_t}$-$\ref{fig:pomegaomega0}), we suggest:

\begin{itemize}
  
\item 
young stars in the Galactic center are formed in one common gaseous disk with marginal 
gravitational stability from $\sim 0.01$ pc to $\sim 0.2$ pc, and possibly beyond;

\item 
the asymptotic mass of the newly formed stars in the inner regions of the disk may be 
thermally limited (such that their Roche radius is comparable to or smaller than their 
natal disk's scale height \citep{li2021}) and the upper limit of their $m_\star$ increases 
with distances from Sgr A$^\star$ in flaring disks;

\item
stars form in the present-day domain of the S-stars acquire modest $e_\star$ (up to $\sim 0.5$)
through resonant relaxation among themselves;

\item
the gaseous and stellar disks are perturbed by an IMC with comparable mass
($\sim 10^4 M_\odot$) and on an inclined (by $120^{\circ}$) orbit;

\item
in the middle region between the S-stars and the IMC, a fraction of
the young stars are weakly perturbed by the IMC and their $e_\star$ and $i_\star$ 
are modestly modified from their initial values;

\item 
the IMC induces large-$e_\star$ and high-$i_\star$ excitation within 3 Myr
(Figure \ref{fig:a_inc_t}) for stars in its proximity and these stars acquire 
less mass from the gaseous disk since they spend only a fraction of their orbital 
period passing through its mid-plane;

\item
in the inner regions of the disk (where the S-stars reside today), resonant relaxation 
between locally formed stars and some intruding stars (with high-$e_\star$ and $i_\star$) 
from the outer region of the disk (where CWSs reside today) rapidly (within $\sim 5-7$ Myr)
leads to phase-mixing and nearly isotropic $e_\star-i_\star$ distribution; and

\item
many stars remain in the thickened outer region of the disk with 
large $e_\star$ and $i_\star$, randomly oriented longitude of ascending nodes 
such that they constitute a marginally counter-rotating torus which may appear 
to be composed of a clumpy substructure including a counter-clockwise disk,
in the distribution of their orbital angular momentum vectors.  

\end{itemize}

%%%%%%%%%%%%%%%%%%%%%%%%%%%%%%%%%%%%%%%%%%%%%%%%%%%%%%%%%%%%%%%%%
\subsection{Predictions and outstanding issues}

With the addition of an IMC, the unified model provides a comprehensive, self-consistent, 
common-origin explanation for the observed complex multi-component kinematic and spatial 
structure of the S-stars, CWSs, and ODSs around Sgr A$^\star$. The nature and boundedness of 
the IMC-candidate, IRS13E, could possibly be explored in details with additional measurements of 
its internal velocity dispersion and mass-to-light ratio, including the question whether it 
hosts an intermediate-mass black hole \citep{schodel2005}. 
High-precision proper-motion observations might in the 
future also provide information about its orbit which can be compared with our predictions
%(Figure \ref{fig:cws_ods}) 
for a single or multiple co-orbital IMCs. 
Eventually one might be able to detect S-stars
orbital precession,
with associated eccentricity-inclination
($e_\star-i_\star^\prime$)
changes \citep{zheng+2020}, which should be induced by the 
IMC's gravitational perturbation on nearby stars, in excess to that due to Sgr A$^\star$'s 
post-Newtonian effects during individual close encounters \citep{heissel2022}.

The above speculative IMC scenario can be tested with further observations. 
We plan to explore this possibility in the presence or absence of the IMC in a follow-up investigation.

\noindent
1) The intriguing prospect of IRS-13E being an 
intermediate-mass ($\sim 10^4 M_\odot$) 
black hole is not required since
IMC's secular distant perturbation (rather than impulsive 
close encounters) can also be provided by a compact (with a 
high mass-to-light ratio) cluster or a moving stream of 
stars ({\it two-IMC} model in Figures \ref{fig:e_peri_a0_t9_mc},
\ref{fig:a_inc_e_t9_mc}, \& \ref{fig:e_peri_a0_t9_IMC}).  Confirmation of IRS-13E being 
a genuine entity bound by its self-gravity or 
a moving-group member would be adequate.  

%\noindent
%2) Although the IMC-CWSs inclination 
%($i_{\rm IMC} ^\prime \sim 60-120 ^{\circ}$,~\S
%\ref{sec:imcinclination}) in the vZLK-effective 
%models is within GAIA's precision limit,  
%IRS-13E's overcrowded field poses challenges. 
%Recent observation \citep{jia2023} of 
%kinematic similarity between the IRS-13 group and 
%the ODSs is consistent with {\it Fiducial} model's 
%orbit-normal distribution (Figure \ref{fig:cws_ods}). 
%Long-term astrometry would improve $i_{\rm IMC} ^\prime$-measurement accuracy.

\noindent
2) This scenario requires a limited range of orbital parameters for the IMC (\S\ref{sec:imcinclination}). 
Based on these constraints, we estimate its transverse motion in the sky to be a few mas yr$^{-1}$
with an inclination $\sim 60-120^{\circ}$ relative to the CWS disk plane (\S\ref{sec:imcinclination}) in the vZLK-effective models. 
%Although this proper motion is within GAIA's precision limit, the field near IRS-13E may be too crowded for such an attempt.
Recent observation data \citep{jia2023} provide tantalizing hints that the IRS-13 group may have similar
kinematic properties as the ODSs. Confirmation of this correlation with future high-precision proper motion
measurements would provide support for the prediction (Figure \ref{fig:cws_ods}) that IMC has a
similar orbit as those of some ODSs.
 
\noindent
3) During the $e_\star-i_\star$ excitation, IMC's vZLK and SSR lead to 
clustered $\varpi_\star$ and diffuse $\Omega_\star$ distribution 
(\S\ref{sec:orientation}). Their precisely observed values can provide 
more stringent constraints on the kinematic distribution of the pseudo counter-clockwise stars and
their surrounded ODS toroid. 
%%% added by zxc

%%% end

\noindent
4) The vZLK, SSR, and resonant relaxation-induced precession 
dominates the post-Newtonian effects on S-stars with $r_{\rm p} \geq 
10^{-2}$ pc (Figure \ref{fig:rp_timescale}).  Moreover, its direction is 
generally off the orbital plane (nodal precession) 
over multiple periods, departing from the Schwarzschild 
apsidal precession during individual close encounters
\citep{heissel2022}.

\noindent
5) Compositional and luminosity-function comparisons between 
the S-stars, CWSs, and ODSs and hyper-velocity stars can test 
their common-origin scenario \citep{zheng+2021} and provide
clues on whether S-stars and ODSs' $M_\star$-distribution 
may be affected by their rapid $i_\star$ excitation due 
to IMC's vZLK and stars' resonant relaxation (\S\ref{sec:mechanisms}).

The present study highlights the relevance of complementarity between composite IMC's secular perturbation
and stars' dynamical relaxation.  Similar processes may play a role in the context of galaxy mergers where multiple embedded SMBHs, with surrounding stars including binaries
\citep{lu2007, zheng+2021}, undergo orbital decay towards coalescence.  
It may also be relevant in the formation of multiple protostellar disks \citep{owen2023} and
planetary systems around embedded stars or ``stellar-host-free''
planetary companions in the proximity of SMBHs.
We will explore these possibilities in future investigations.

%%%%%%%%%%%%%%%%%%%%%%%%%%%%%%%%%%%%%%%%%%%%%%%%%%%%%%%%%%%%%%%%%%%%%%%%
%\subsubsection{Some outstanding technical issues and uncertainties}
%\label{sec:unsolved}

Here, we assume a population of coeval, single,
identical-$M_\star$ stars, and neglect their evolution 
and dynamical relaxation with individual low-mass 
mature, main-sequence stars in the neighborhood. But, 
Sgr A$^\star$ is surrounded by a cluster of old stars 
with a non-negligible contribution to the local mass 
distribution and gravitational potential \citep{schodel2007, 
trippe2008, lockmannetal2009}.  They may slightly modify the stars' 
precession frequency \citep{kocsis2015}, lead IMC to undergo limited orbital
decay through dynamical friction \citep{gerhard2001, Hansen2003, Gurkan2005,  
Merritt2009}, potentially excite resonant friction to disrupt
a disk of young stars \citep{levin2024}, and provide seeds 
for stellar trapping \citep{artymowicz1993, davies2020, wang2024binary}.
These effects can be simulated with the PeTar code \citep{wang+2020b} 
in follow-up investigations.

We only consider the gas contribution to the 
axisymmetric gravitational potential, and neglect 
relativistic corrections in most comparisons 
(except the {\it pN} model) models.  Protracted star formation 
\citep{Goodman2003, thompson2005},
capture and rejuvenation \citep{artymowicz1993, davies2020},
gas accretion \citep{cantiello2021, alidib2023, li2021},
stellar wind, tidal interaction with the disk 
\citep{macleod2020, wang2024binary}, type 1 \citep{Perets2009} 
and stochastic \citep{wuchenlin2024} migration may modify stars' 
$M_\star$-distribution and their dynamical evolution \citep{zhou2007}.   
IMC's perturbation to the 3D disk structure may also affect its as well as nearby
stars' precession rates and therefore modify the impact of its vZLK
and SSR effects. 
Simulation of the combined effect of hydrodynamic and \textit{N}-body interaction requires the construction of high-resolution, 3-dimensional hybrid codes, which are beyond the scope of this paper.
The current study provides a motivation for the construction of such advanced 
computational tools.

Although we have shown both the hyper-velocity stars \citep{zheng+2021} and 
the observed $r_{\rm p}$-$(1-e_\star)$ correlation arise naturally {\it on the time scale
comparable to the a-few-Myr stellar age} from IMC's perturbation on single stars, 
without invoking any tidal-disruption-of-binary hypothesis, massive stars, similar to those around Sgr A$^\star$ are often members of binary systems.
The S-stars' periastra are sufficiently far away for them 
to avoid tidal disruption 
\citep{Hills1975, frank1976}. The rarity of binary S-stars 
\citep{chu2023} remains an outstanding issue since most 
field massive stars have companions and binary capture in 
AGN disks may be common \citep{wang2024binary}. If this discrepancy is 
due to binary's tidal breakup \citep{Hills1988, Perets2009}, it
would also enhance the S-stars' velocity-dispersion growth,
stellar disintegration, or ejection of hyper-velocity stars.  
%%% added by zxc
The predicted distribution of orbital planes for S-stars is not entirely isotropic, and the possible reasons can be attributed to incomplete relaxation, initial conditions, or the need for additional mechanisms such as the Hills mechanism.
%%% end
Although Sgr A$^\star$'s Lense-Thirring effect 
\citep{lense1918, merritt2013, rodriguez2018, iorio2020} may 
be non-negligible for disrupted or ejected stars 
\citep{zheng+2021}, its precession timescale ($\simeq 
(2 r_{\rm p}/a_\bullet)^{3/2} P_\star/2 \pi \chi 
\geq {\mathcal O} ({\rm Myr})$, where $0<\chi 
\leq 1$ \citep{EHT2022a, EHT2022b} is the spin 
parameter) is $\geq \tau_{\rm RR}, \tau_{\rm vZLK},$ and 
$\tau_{\rm SSR}$ and it does not influence the 
dynamical evolution of most S-stars, CWSs, and ODSs.
%%% added by zxc
Our scenario may operate alongside other processes, rather than in isolation.
%%% end

Future investigation for a range of stellar masses may provide information on the
prospects of mass segregation \citep{spitzer1987} and enhanced resonant relaxation. 
Stellar age spreads over Myrs in a typical dense star-forming complex. Protracted star formation 
\citep{Goodman2003, thompson2005} or capture and rejuvenation \citep{artymowicz1993, davies2020} over the
active phase of Sgr A$^\star$ may diversify some of the outcomes we have represented here.
Before their netal disk is severely depleted, the newly formed or captured stars
can continue to gain mass through accretion \citep{cantiello2021}.
We suggest that the CWSs are more massive than the S-stars due to 
these stars' thermally-limited asymptotic mass (with stars' Roche radius being 
comparable to the disk's scale height \citep{li2021}) being an increasing function
of the disk radius. We also proposed that growth of ODSs is quenched by the
rapid excitation of their $i_\star$.  These hypotheses need to be verified with ongoing 3D hydrodynamic
simulations of tidal interaction between multiple stars and their natal disk.
In addition, massive stars lose mass and undergo SN II, which may introduce 
additional dynamical perturbation to the young stars' dynamical evolution \citep{hills1980, wang2021}. 

In our future work, we will also explore an additional piece of evidence that supports this unified scenario: the observed downturn in the number density of late-type (Bahcall-Wolf) stars. This feature is challenging to explain with standard cluster relaxation models alone. Our scenario may provide a natural mechanism that can efficiently heat and scatter the pre-existing, older stellar population. This impulsive excitation between eccentric intruders and the inner stellar population would deplete the central cusp of stars, creating a localized dip in the radial number distribution that aligns with the most recent observation.

%%%%%%%%%%%%%%%%%%%%%%%%%%%%%%%%%%%%%%%%%%%%%%
\section*{Acknowledgments}

We thank Drs Gongjie Li, Jessica Lu, and Hangci Du for the useful discussions.  The authors acknowledge the Tsinghua Astrophysics High-Performance Computing platform at Tsinghua University for providing computational and data storage resources that have contributed to the results in this paper.  This work is partly supported by the National Science Foundation of China (Grant No. 11821303 and 12133005) and Talents Program (24CE-YS-08). XCZ is supported by the National Natural Science Foundation of China (Grant No.12203007) and the Mengya Program of Beijing Academy of Science and Technology (BGS202203). AB's research is supported by the Excellence Cluster ORIGINS which is funded by the Deutsche Forschungsgemeinschaft (DFG, German Research Foundation) under Germany's Excellence Strategy - EXC-2094-390783311. L.W. thanks the support from the one-hundred-talent project of Sun Yat-sen University, the Fundamental Research Funds for the Central Universities (22hytd09), Sun Yat-sen University and the National Natural Science Foundation of China through grant 12073090 and 12233013. This work is supported by the China Manned Space Program
with grant no.CMS-CSST-2025-A16.
%\end{acknowledgments}

\software{PeTar \citep{wang+2020b}}

\bibliography{s-star}{}
\bibliographystyle{aasjournal}



\end{CJK*}
\end{document}